\documentclass{article}%
\usepackage{amsmath}
\usepackage{graphicx}
\usepackage{amsfonts}
\usepackage{amssymb}%
\providecommand{\U}[1]{\protect\rule{.1in}{.1in}}
\newtheorem{theorem}{Theorem}
\newtheorem{acknowledgement}[theorem]{Acknowledgement}

\begin{document}

\title{String theory and the crisis in particle physics II \\or the ascent of metaphoric arguments \\{\small Dedicated to the memory of Juergen Ehlers}\\{\tiny (invited contribution to IJMPD)}}
\author{Bert Schroer\\CBPF, Rua Dr. Xavier Sigaud 150 \\22290-180 Rio de Janeiro, Brazil\\and Institut fuer Theoretische Physik der FU Berlin, Germany}
\date{May 2008}
\maketitle
\tableofcontents

\begin{abstract}
This is a completely reformulated presentation of a previous paper with the
same title; this time with a much stronger emphasis on conceptual aspects of
string theory and a detailed review of its already more than four decades
lasting history within a broader context, including some little-known details.
Although there have been several books and essays on the sociological impact
and its philosophical implications, there is yet no serious attempt to
scrutinize its claims about particle physics using the powerful conceptual
arsenal of contemporary local quantum physics.

I decided to leave the previous first version on the arXiv because it may be
interesting to the reader to notice the change of viewpoint and the reason
behind it.

Other reasons for preventing my first version to go into print and to rewrite
it in such a way that its content complies with my different actual viewpoint
can be found at the end of the article.

The central message, contained in sections 5 and 6, is that string theory is
not what string theorists think and claim it is. The widespread acceptance of
a theory whose interpretation has been obtained by metaphoric reasoning had a
corroding influence on the rest of particle physics theory as will be
illustrated in several concrete cases.

The work is dedicated to the memory of Juergen Ehlers with whom I shared many
critical ideas, but their formulation in this essay is fully within my
responsibility. 

\end{abstract}

\section{An anthology of the crisis in the foundations of particle physics}

There can be no doubt that after almost a century of impressive success
fundamental physics is in the midst of a deep crisis. Its epicenter is in
particle theory, but its repercussions may influence the direction of
experimental particle physics and affect adjacent areas of fundamental
research. They also led to quite bizarre ideas in the philosophy of
fundamental sciences, which partially explains why they attracted much general
interests beyond the community of specialists in particle physics.

One does not have to be a physicist in order to be amazed when reputable
members of the particle physics community \cite{Suss} recommend a paradigmatic
change away from the observational based setting of physics which, since the
time of Galileio, Newton, Einstein and the protagonists of quantum theory and
quantum field theory has been the de-mystification of nature by mathematically
formulated concepts with experimentally verifiable consequences. The new
message which has been formed under the strong influence of string theory is
that it is scientifically acceptable to use ones own existence in reasonings
about theoretical physics matter even if this leads to a vast collection of in
principle unobservable concepts such as multiverses and parallel worlds. This
new physics excepts metaphors (but calls them princples as e.g. the
\textit{anthropic principle}); its underlying philosophy resembles religion
with its unobservable regions of heaven and hell rather than physics as we
know it since the times of Galileio. It certainly amounts to a rupture with
natural sciences and with the accompanying philosophy of enlightenment.
despite all assurances to the contrary it looks like a parallel universe
avatar of \textit{intelligent design}.

Instead of \textquotedblleft cogito ergo sum\textquotedblright\ of the
rationalists, the new \textit{anthropic} maxim coming from this new doctrine
also attaches explanatory power to its inversion: I exist and therefore things
are the way they are since otherwise I would not exist. Its main purpose is to
uphold the uniqueness of the string theorists dream of a TOE. Even with an
enormous number of different solutions with different fundamental laws and
fundamental constants one is able to claim that these describe actually
existing but inaccessible parallel worlds or multiveres of a unique multiverse
TOE. Anthropic reasoning here is not meant as a temporary auxiliary selective
device, pending better understanding and further refinements of the theory,
but rather as the endpoint of a theory. The logic behind this doctrin vaguely
resembles the "if you cannot solve a problem then enlarge it" motto of some politicians.

The invocation of the anthropic principle serves the purpose to uphold the
holy Grail of string theory as a TOE. To demonstrate the physical relevance of
string theory it would suffice to show that there is one solution which looks
like our universe; but whereas the number of solutions has been estimated,
nobody has an idea how to arrange such a search. How can one find something in
a haystack if one does not even know how to characterize our universe in terms
of moduli and other string-theoretic notions which fix what string theorists
call the vacuum?

In his vein one could even propose the idea that all possible QFT models taken
together form a unique TOE called the TOE of local quantum physics. In this
case the underlying principles (Poincar\'{e} covariance, causal locality and
positive energy) are even known. By this semantic trick we can claim that we
have been dealing with a TOE for almost one century without having been fully
aware of it\footnote{This is in fact the point of view in \cite{Teg} where an
even more extreme concept of physical reality is proposed which consists in
claiming mathematically consistent theory attains a physical reality in at
least one universe of the multiverse.}.

To be fair, the anthropic dogma of a \textit{multiverse} instead of a universe
i.e. the belief that all these different solutions with quantum matter obeying
different laws (including different values of fundamental constants) exist and
form "the landscape" \cite{Suss}, is not shared by all string theorists. \ 

Although such a picture is still confined to a vocal and politically
unfluential minority in particle theory, it is not difficult to notice a
general trend of moving away from the traditional scientific setting based on
autonomous physical principles towards more free-wheeling metaphoric
consistency arguments. It seems that as some physicists are moving away from
religion, theology takes its payback by the increase of metaphoric physical
arguments. Ironically the new agressive science-based atheists are strong
defenders of the metaphors about the string-inspired multiverse.

The ascend of this metaphoric approach is strongly supported by the increasing
popularity of string theory and the marketing skill of its proponents to
secure a lavish funding. This, and certainly not the extremely meager physical
results, is what at least partially accounts for its present hegemony status
in particle theory. Whereas the attraction it exerts on newcomers in physics
is often its career-accelerating quality, the broader scientifically
interested public finds the media hype, which highlights its "revolutionary"
achievements highly entertaining. There are also more general sociological
reasons which even in the absence of a TOE quality aggravate scientific
progress, some of them will be mentioned in the last two sections.

Parallel to this development, the particle physics community experienced a
deepening frustration as a result of inconclusive or failed attempts to make
further progress with the standard model. This model has remained particle
physics finest achievement ever since its discovery more than three decades
ago. It continued the line of moving towards unification which started already
with Faraday, Maxwell, Einstein Heisenberg and others. This kind of
unification has been the result of a natural process of the development of
ideas, i.e. the protagonists did not set out with the aim to construct a TOE,
rather the unification was the end result of a natural unfolding of ideas
following the intrinsic theoretical logic but with observations and
experiments having the ultimate power to decide whether a mathematically
consistent theory is also realized in nature. The relation of this
old-fashioned unification to the modern TOE is not unlike that of the old
style capitalism to its unleashed globalized counterpart. Whereas the old
version still showed social responsibility, the new one is mainly about
hegemony i.e. power and glory. There and here ethical erosions have left their mark.

A previous IJMPD-invited version of this article was stopped at proofreading.
One reason is that I got disenchanted with my style of criticism which often
was as metaphoric as the subject which it aimed to criticize. As a result the
present version became more mathematical and conceptual. I also was able to
enrich the historical part with additional facts. Additional more specific and
personal reasons will be explained in the last section.

The detailed mathematical content of my critique of string theory is very
different from that of high energy experimentalists as Burton Richter who
among other things $_{{}}$calls supersymmetry a "social construct". It is also
different from that of condensed matter theoreticians as Phillip Anderson and
Robert Laughlin who use their rich professional experience in their own very
successful area in order to criticize string theory on its total lack of
observable predictions and its almost new age like metaphoric way of arguing.
Although I agree with their conclusions I will not repeat their arguments in
this essay.

I think a particle theoretician should take his criticism right into the
conceptual-mathematical core. Instead of "not even wrong" (which has a
metaphorical connotation) one should aim for \textit{even wrong}. The
strongest concrete anti-string argument is that string theorists understanding
of string-localization (which led to the name) is really a complete
misunderstanding. This will be the main theme in section 4.

Of course a scientific critique cannot answer the question why so many people
permit themselves to be led by a TOE into a profound conceptual glitch. This
will be left to sociologists and historians of physics who, as a result of the
magnitude of the problem, will certainly attract a very interested public who
will be eager to understand what happened to the millennium TOE.

Particle physics is a conceptually and mathematically quite demanding science
and its progress depends on the one hand on a\ "into the blue yonder" spirit,
but on the other hand, in order to go not astray, it needs the delicate
balance with a cutting criticism whose intellectual depth at least matches
that of its object. To some extend this is a pure inner theoretical process in
which the main issue is that of conceptual consistency. Particle theory is
very rich in established fundamental principles and it also has a very strong
time-hardened intrinsic logic. Experiments cannot decide whether a theoretical
proposal is consistent in its own right, but they can support or reject a
theory or select between several consistent theories.

My main thesis is that the conceptual error in string theory occurred right at
the beginning i.e. at the time when the dual model passed into string theory.
Two previous attempts at a pure S-matrix theory, Heisenberg's 1943 proposal
and the Chew-Mandelstam-Stapp bootstrap idea of the 60s were too vague in
order to pinpoint any glitch within them. The dual model/string theory on the
other hand replaced the important crossing property by something which is
ostensibly formally related, namely the Veneziano-Dolen-Horn-Schmid duality,
but whose conceptual status remained unknown. Nevertheless, apart from
unitarity, the DHS dual resonance model was much more concrete than any
previous S-matrix proposal.

At the time of its inception the dual resonance model was considered as a
curiosity by quantum field theorists because despite its roots in Regge
phenomenology it was surprizingly precise and explicit and it represented at
least some desirable and rather intricate properties of the S-matrix.

The conceptually somewhat opaque relation of infinite particle towers
(required by the DHS duality) to the field theoretic
crossing\footnote{Crossing is one of the most subtle analytic properties of an
S-matrix coming from QFT. It relates (a finite number of) particle poles with
the scattering continuum. More can be found in section 4.} of its S-matrix
bootstrap predecessor, would have served as an ideal starting point for a
systematic investigation of the relation of crossing to QFT as well as
crossing to DHS duality. Such a line of research would have possible led to a
better understanding of the spacetime origin of DHS duality and a better
conceptual positioning of the Nambu-Goto model and string theory.

The rather short interim between the (at that time still fameless) dual model
and the later glorious string theory could have served a window for criticism
of string theory with a good chance to be listened to, since there were yet no
big fan-club which usually blur the critical vision. But probably lack of
physical motivation and mathematical attraction for those particle physicists
who had the capabilities to investigate such hard conceptional/mathematical
problems contributed to this missed opportunity. In those days string theory
was not yet that strong protective bulwark for a large community hardened by
several revolutions.

The present view of string theorists at their own history is reflected in the
several recent 40 year anniversary contributions, some from veterans of the
dual model days. They make quite interesting historical reading; but whoever
looks for critical comments on the strange contradiction between physical
misery and social glamour of string theory will be disappointed; even after 40
years the time for a critical evaluation has not yet arrived.

In line with our viewpoint that every speculative idea in particle physics is
welcome as long as the balancing critical counterpart is in place, there is no
intention to blame persons for ideas which later led to dominating fashions
which not only did not live up to their promises but also contaminated
particle theory with metaphoric arguments. It is not the protagonists of such
ideas but rather \textit{our failure to subject them to profound conceptual
scrutiny} at the right time which derails particle physics..

In the case at hand there was no critical review; not because people in the
70s we are less intelligent than their predecessors, but rather because the
TOE based on superstrings dominated the particle physics scene quite rapidly
and the leading established and dominating figures, who in earlier times would
have taken a critical look at new ideas, became the strongest defenders of a
new TOE (presumably as a result of its promise to include quantum gravity). In
this way a problem which started in particle physics finally led to the idea
of a TOE and became part of the millennium Zeitgeist of power and glory,
weakening the conceptual basis of the traditional particle physics approach
and pushing it finally into the sidelines.

I have no problem to admire people as Gabrielle Veneziano and the other
protagonists of the dual model, even if on the other hand I am convinced that
string theory is the first mayor derailment in particle physics. I hope this
makes it possible for string theorists to look also at the present scientific
criticism with a certain emotional detachment and rational attention.

This intention to go to the conceptual roots also separates the content of the
present reworked presentation from its previous version, as well as from the
various string-critical books, articles and published statements by particle
physicist, philosophers and condensed matter physicists as P. Woit \cite{Wo},
L. Smolin \cite{Sm} and R. Hedrich \cite{Hedrich}. In those articles the
consistency of its conceptual-mathematical framework was not the issue; their
content is not directed to test the mathematical-conceptual consistency of
string theory but their main concern is the lack of tangible results despite
of more than four decades of work by hundreds of brilliant minds and,
particularly in case of Lee Smolin, the resulting futile consummation of
valuable resources\footnote{There remains the question however why a permanent
member of the Perimeter Institute does not use his influence to create a space
for critical discussions about what string theory really is, as opposed to
what its supporters think it is.}. The critical comments of philosophical
adversaries of string theory tend to be directed towards its metaphorical
anti-Popperian way and its rupture with Heisenberg's principle of limiting
arguments to observables which he and his contemporaries recognized to be
absolutely crucial in order to avoid classical arguments contaminating quantum
theoretical arguments.

What is somewhat surprising is the conspicuous absence of any profound
critique coming from particle theorists, especially from mathematical
physicists. A theory with no predictive power could still be consistent, but
if it comes with has a permanent conceptual flaw it should be dismissed; in
that case there is nothing which can be learned, even if by luck or
coincidence, it "explains" some facts\footnote{The reason why the phlogiston
theory was able to hold on for some times was that its predictions actually
agreed with several observed facts up to Lavoisier's crucial experiment which
brought its demise.}.

The content is structured as follows.

The next section reviews Heisenberg's S-matrix proposal and Stueckelberg's
profound criticism on the basis of its macro-causality defects (which led him
to the discovery of Feynman rules several years before Feynman). The third
section recalls the S-matrix bootstrap program whose lasting merit consists in
having added the important on-shell crossing symmetry to the requirements of
an S-matrix program. The fourth section analyses the origin of on-shell
crossing from off-shell localization concepts and comments on its proximity to
the KMS thermal aspect of localization. Section 5 reviews the implementation
of duality of the DHS dual resonance model in the setting of charge
superselection property of a multi-charge chiral current model with is
intrinsic to chiral conformal theories. In this way the differences between
duality resulting from the plektonic commutation relations of charge-carrying
chiral fields and the particle-based notion of crossing becomes highlighted.
The previous results on quantum localization obtained in the fourth section
are then used in section 6 to show that string theory contrary to its claims
does not deal with string-localized objects in spacetime; its simplest
(interaction-free) realization looses its classical string-like appearance
under canonical quantization and its associated string fields is pointlike
localized but with many more degrees of freedom than those coming from a field
theoretic Lagrangian; technically speaking it is a generalized free field with
an infinite mass and spin spectrum.

The last two sections attempt to shed some light on how it is possible that a
theory with so many conceptual shortcomings and glitches (extending right up
to the quantum physical meaning of its name) is selected by a worldwide
community of particle physicists to represent the power and glory of the
millennium particle physics. So in those sections we leave the ivory tower of
particle physics and turn to the millennium Zeitgeist.

Since the mathematical-conceptual content is quite demanding, some statements
and arguments will appear more than once in a different formulation and
context. It is hoped that using this essay style of shedding light on one
aspect from slightly different angles will make the main arguments more accessible.

\section{QFT versus a pure S-matrix approach from a historical perspective}

Particle physics was, apart from a 10 year period of doubts and confusion
around the ultraviolet catastrophe starting in the late 30s, a continuous
success story starting from its inception \cite{Dar} by Pascual Jordan
(quantization of wave fields for light and matter) and Paul Dirac
(relativistic particles and anti-particles via hole theory) up to the
discovery of the standard model. For about 40 years the original setting of
Lagrangian quantization, in terms of which QFT was discovered, gave an ever
increasing wealth of results without requiring any change of the underlying
principles, apart from some conceptual and mathematical refinements in order
to adjust the formalism of QT to causal propagation with the velocity of light
as the limiting velocity. After the clouds of doubts about the ultraviolet
catastrophe dissolved, thanks to the new setting of covariantly formulated
perturbative renormalization theory, the conceptional and mathematical
improvements reinforced the original principles.

It is interesting to observe that already at the beginning of QFT even its
protagonist Pascual Jordan worried about the range of validity of
quantization. These doubts originated from his conviction that, although
classical analogies allow in many cases rapid access to the new quantum theory
of fields in form of important illustrations, in the long run a more
fundamental quantum theory should not need the parallelism to the less
fundamental classical Lagrangian formalism referred to as quantization, but
rather develop its intrinsic arsenal of classification and construction of
QFTs, or in his words "without borrowing crutches" from the less fundamental
classical theory \cite{Kha}. To turn the argument around: to the extend to
which one has to rely on quantization crutches, one has not really reached the
core of the new theory.

Jordan's doubts about the range of validity of that umbilical cord to
classical field theory did not originate from any perceived concrete
shortcoming of his "quantum theory of wave fields". Rather the state of
affairs in which he discovered this new theory did not comply with his
philosophical senses; in his opinion this can only be tolerated as a temporary
device for a quick first exploration of those parts of the new theory which
are in the range of this quantization recipe.

But things did not develop in the direction of Jordan's plea. The ultraviolet
divergence crisis of the 30s ended in the late 40s in the discovery of
renormalized QED, a fact which certainly revitalized the Lagrangian approach
and pushed the search for an intrinsic formulation into the sideline.

Unfortunately the renormalized perturbation series of quantum field
theoretical models diverges, so the hope to settle also the existence problem
of QFTs in the Lagrangian quantization setting did not materialize; the
success of the renormalized perturbative setting did not lead to a conceptual
closure of QFT. However at least it became clear that the old problem of
ultraviolet infinities, which almost derailed the development of QFT, was in
part a pseudo-problem caused by the unreflective use of quantum mechanical
operator techniques for pointlike quantum fields which are too singular to
qualify as operators.

Using more adequate mathematical tools in conjunction with a minimality
principle which limits the short distance singularity of the undetermined
total diagonal contribution in terms of the scaling degree of the uniquely
determined non-diagonal part \cite{E-G}, one finds that there are local
couplings between pointlike fields for which the perturbative iteration either
does not require more parameters than there were in the beginning, or adds
only a finite number of new couplings which one could have already included in
the starting interaction expressed in terms of Wick-product of free
fields\footnote{I am referring here to the Epstein Glaser \cite{E-G}
formulation which produces the renormalized finite result directly by treating
the fields in every order according correctly according to their singular
nature. The avoidance of intermediate cutoffs or regularizations maintains the
connection with the quantum theoretical Hilbert space structure of QFT.}. The
renormalized theory forms a finite parameter space on which the
(Petermann-Stueckelberg) renormalization group acts ergodically. These finite
parametric families are conveniently pictured as "islands" in an infinite
parameter setting (the Bogoliubov spacetime dependent operator-valued
S-functional or the Wilson universal renormalization group setting) within the
"sea" of infinitely many coupling functions (which by itself has no predictive
power). Since the renormalization group leads from any point on the island in
coupling space to any other such point, a QFT cannot provide a method to
distinguish special numerical values. \ 

The phenomenon of interaction-caused infinite vacuum polarization clouds
(finite in every order perturbation theory) gives rise to a conceptual rupture
with QM \cite{interface} and leads to a change of parameters in every order.
But since these parameters remain undetermined anyhow, this causes no harm.
The inexorable presence of interaction-induced vacuum polarization simply
prevent one to think of an initial numerical (Lagrangian) value for these
parameters which is then changed by a computable finite amount. With other
words unlike in QM there is no \ separable "bare" and "induced" part which
only lead to finite values by compensation between them. This is why the
Epstein-Glaser renormalization is conceptually preferable \cite{E-G}. It not
only addresses the singular nature of fields but it also exposes the limits of
QFT concerning the predictive power about the numerical value of certain
parameters in a more honest way. \ 

So when string theorists say that their theory is ultraviolet finite whereas
QFT is not, what they really mean in intrinsic terms is their theory is more
economical (and hence more fundamental) in that it has only the parameters
which describe string interactions i.e. the string tension. But beware, they
say that without being able to give a proof.

This implies that in particular that \textit{string theory has no vacuum
polarization} which is of course completely consistent with its on-shell
S-matrix character. An S-matrix is the object par excellence without vacuum
polarization; in fact Heisenberg's plea for basing particle physics on the
S-matrix was proposed because the S-matric in contrast to quantum fields is
like QM completely free of vacuum polarization and the ensuing apparent
ultraviolet problems. But can one really do particle physics without such a
central concept as vacuum polarization? By what conceptual trick can an
S-matrix theory emulate vacuum polarization? Is the idea of a natural
off-shell extrapolation without the guidance of QFT self-deception? Is there
really any other way then constructing an S-matrix as the large-time limit of
some quantum theory with some spacetime aspects?

Before starting to criticize string theory, one should however look for
imperfections in one's own backyard which in my case means exposing some weak
points of QFT.

The power counting restriction to dim $\mathcal{L}_{int}=$ $d$ (spacetime
dimension)\footnote{The $\mathcal{L}_{int}$ is only a name for the interaction
in terms of free fields. The causal perturbative approach (in contrast to the
functional integral setting) does not use the Lagrangian formalism but only
this local interaction term whereas e.g. the functional integral approach is
limited to free fields which are of the Euler-Lagrange type.} is quite severe
because for $d=4$ it only allows (without using "ghost crutches") pointlike
fields $\Phi$ with dim$\Phi<1.$ In addition massless vectorpotentials (and
more generally the potentials associated with the Wigner massless finite
helicity representations) \textit{cannot be pointlike covariant objects within
a (ghost-free) Wigner Fock representation}; the best one can do is to describe
them by semiinfinite stringlike-localized fields $A_{\mu}(x,e)$ \cite{MSY}%
\ with the spacelike unit vector $e$ being the string direction (no relation
of string localization to string theory, as will become clear in later
sections). These massless string potentials exists for all higher helicities
and as a result of their non-compact localization they all have scale equal to
one which makes them ideal candidates for fulfilling the renormalizability
criterion in trilinear and quadrilinear interactions. This is not a technical
side remark but the result of the fundamental quantum requirement that the
generalized potentials associated to the massless generalized field
strengths\footnote{For s+2 the field strength is the (linearized) Riemann
tensor and the potential is a string-localized linearized metric tensor
$g_{\mu\nu}(x,e)$ localized along the line $x+\mathbb{R}_{+}e$ (see section
4). The string localization comes from quantum requirements and has no
counterpart in the classical theory.} ought to be objects in physical space

The requirement of finding pointlike localized composite fields imposes a
severe restriction on the interactions which turn out to be equivalent to the
gauge invariance in a gauge theoretic formulation using ghosts. The basic
generating fields are stringlike localized but these fields have composites
which are usual pointlike fields.

As mentioned in case of massive higher spin fields the string localization has
the same short distance improving effect even though there is no argument
necessitating stringlike localized potential for the description of the Wigner
representation space which can be perfectly described in terms of pointlike
field strengths. For the time being the requirement to be able to formulate
renormalizable interactions is the only one but there is also a rather subtle
indication of their presence in the free theory via the violation of Haag
duality for multiply connected causally closed spacetime regions \cite{MSY}.
The relation between the stringlocalized potential and the pointlocalized
field strengths changes drastically in the presence of selfinteracting potentials.

A perturbation approach for stringlike localized representation of free fields
has not yet been formulated. Instead one evades the No-Go theorem from
Wigner's representation theoretical approach by maintaining pointlike
covariant localization and instead sacrificing the Hilbert space setting
through the introduction of (BRST) ghost fields. This \textit{gauge} field
formalism makes helpful contact with classical gauge theory (where the Hilbert
space aspect plays no role) and permits the use of the well-known pointlike
renormalization machinery whose perturbative version does not care about
indefinite metric. A consistent descend to the physical representation in a
ghost-free Hilbert space is guarantied by the cohomological properties of the
BRST ghost formalism. There are however conceptual limitations of this
formalism (in particular with respect to the Higgs issue) which will be
mentioned in a later section.

This short account of the history of QFT and particle physics contains most of
the ideas which are needed for the formulation of the standard model which
places QED, the weak interaction and the QCD setting of strong interactions
under one common gauge theoretic roof. But it also was meant to expose the
gaping unfinished areas of QFT. Anybody who claims that QFT is a closed
subject and that its innovative role has passed to string theory does not know
what he is talking about. The unification which led to the standard model is
natural i.e. the desire to obtain a TOE was not the motivation. Whether the
running coupling constants of the three interactions really come together at a
sufficiently high energy and whether gravitation can be incorporated remain
open questions outside the standard model.

One of the marvelous achievements of the post QED renormalization theory is a
clear understanding of the particle-field relation (not to be confused with
the particle-wave dualism in QM) in the presence of interactions. Whereas in
free field theories Heisenberg observed the presence of vacuum fluctuations
due to particle-anti-particle pairs in states obtained by the application of
(Wick) composites to the vacuum, the real surprise came when Furry and
Oppenheimer discovered that in interacting theories even the Lagrangian field
generates vacuum polarization upon application to the vacuum state. Different
from the case studied by Heisenberg, the interaction-caused pairs increase in
number with the perturbative order and form a \textit{vacuum polarization
cloud} containing an infinite number of virtual particles. This observation
challenges the naive identification of particles and \ fields which is the
result of a simple-minded conceptual identification of QFT as a kind of
relativistic QM. Although one-particle states exist in the Hilbert space and
the global operator algebra certainly contains particle creation/annihilation
operators, \textit{compactly localized subalgebras\footnote{The only
localization which allows PFGs is the non-compact wedge-like localization
\cite{BBS}.}} in interacting QFTs contain no vacuum-\textbf{p}%
olarization-\textbf{f}ree \textbf{g}enerator (PFGs) i.e. no operator which
creates a one-particle state from the vacuum without contamination from vacuum polarization.

The particle field relation was partially unveiled when in the post QED
renormalization period it became clear that apart from one-particle states
\textit{QFT is not capable to describe interacting particles at a finite
time}; as a result of the ubiquitous presence of vacuum polarization clouds it
is only possible to have an asymptotic description when, barring long range
forces and infrared problems, the localization centers of particle are far
removed from each other so that the interaction does not matter. In fact the
elaboration of scattering theory as a structural consequence of causal
locality, energy-momentum positivity and the presence of a mass gap in the
late 50s and early 60s was one of the finest conceptual achievements of
particle theory.

As mentioned in the previous section, the idea of a pure S-matrix theory as a
remedy against the ultraviolet catastrophe of the old (pre-renormalization)
QFT was first proposed by Heisenberg\footnote{The concept of a unitary
scattering operator as a mapping incoming multiparticle configurations into
outgoing in the limit of infinite timelike separations was introduced
independently by Wheeler and Heisenberg. however the idea of a pure S-matrix
theory as an antidote against the pre-renormalization pretended ultraviolet
catastrophe is attributed to Heisenberg.} \cite{1946}. The S-matrix models
with which he illustrates his paper resulted from a naive unitarization of the
interaction Lagrangian (see next section). Heisenberg's proposal was
immediately criticized by Stueckelberg who pointed out that, although it was
Poincar\'{e}-invariant and unitary, it did not meet the requirements of
macro-causality (see next section).

In the next section we will use Heisenberg's construction and isolate the
problem on which every pure S-matrix theory failed: fitting together unitarity
and Poincar\'{e} covariance with macrocausality (notably the cluster
factorization property). Clustering is the spacelike aspect of macro-causality
which is indispensable for any S-matrix whether its comes from QFT or any
other theory of interacting particles. In QFT and other off-shell
implementations of particle interactions, the clustering property is
implemented on correlation functions or (similar to nonrelativistic QM)
through asymptotic additivity of the interaction-dependent generators of the
Poincar\'{e} group. Its validity for the asymptotic configurations is then a
side result of the proof of asymptotic convergence. With other words, the
highly nonlinear on-shell unitarity requirement is trivialized by showing that
it results from the large time limiting of more easily implementable linear
additive clustering properties for correlation functions.

A long time after the Heisenberg proposal went into oblivion and QFT
experienced a strong return in the form of renormalized quantum
electrodynamic, ideas about the S-matrix returned again, this time as the
result of the uselessness of perturbative arguments in strong interactions
between mesons and nucleons. They led to the S-matrix bootstrap by Chew and
Mandelstam with some ideological backing by Stapp. There was also a popular
version intermingled with Buddhism by F. Capra which was aimed at the (in
those years very strong) world-wide Hippy community.

The analytic aspects of QFT correlations, which follow from locality and
spectral properties, imply an attribute which was first seen in Feynman
diagrams within a fixed perturbative order. Restricting the external legs of
these graphs to the mass-shell in order to obtain perturbative contribution to
the S-matrix, one could show that the different S-matrix elements belonging to
different distributions of n-particles into $k$ incoming and $l$ outgoing
particles are connected by an analytic continuation. The surprizing aspect
(which was not trivial even with Feynman graphs) was that this was possible
without leaving the complexified mass shell. With other words crossing is not
a symmetry but rather an analytic on-shell mark left by the spacelike
commutativity of QFT. Although there is no general proof of crossing for
generic particle configuration, most particle physicists would agree that
highlighting this property will remain as one of the few legacies of that
bootstrap period.

The bootstrap community never exhibited a model in which this new property is
nonperturbatively realized, in fact the concept of a model hardly makes sense
in the bootstrap setting of a TOE, it is either everything or nothing. On the
other hand, as in Heisenberg's first approach, macro-causality and in
particular the cluster factorization of S was not even mentioned; one can
safely assume that after its importance was pointed out in the work of
Stueckelberg (shortly after Heisenberg) it was again forgotten.

There exists an exceptional situation in d=1+1 which is related to the
kinematical equality of the energy-momentum delta function with the product of
two one-particle delta functions. This has the effect that the cluster
property cannot separate the 2-particle interacting contribution from the
identity term of S. In this case it is possible to classify pure 2-particle
elastic S-matrices and represent the n-particle amplitude in terms of the
two-particle amplitude through a combinatorial formula \cite{Ba-Ka} which is
compatible with the d=1+1 cluster decomposition. Purely elastic relativistic
scattering in higher spacetime dimension is only possible in the relativistic
quantum mechanics of direct particle interactions \cite{Co-Po} but not in QFT.

\section{Unitarity and macro-causality in relativistic particle theories}

There are three fundamental requirements which every S-matrix of relativistic
particle physics must obey (and there is no dispense for string theory which
claims to be as an S-matrix theory of relativistic particles) namely:
Poincar\'{e} invariance, unitarity and macro-causality. None of these concepts
requires to introduce fields; macrocausality is a very weak version of
causality which can be formulated and understood in terms of only particle
concepts. To avoid misunderstandings, there are analytic properties of
scattering amplitudes as, e.g. the crossing property, for which the necessary
analytic continuation takes place inside the complex mass shell; but such
on-shell properties cannot be traced back to principles referring to particles
only. Rather they must be understood as being an on-shell imprint of the
causal locality principles of an underlying local quantum physics i.e. the
on-shell particle objects are only the projection screen for analytic
manifestations which originate from off-shell causal locality properties.

As a pedagogical exercise which leads us right into the problematic aspects of
pure S-matrix theories let us revisit the situation at the time when
Stueckelberg criticized Heisenberg's S-matrix proposal.

As a way out of the ultraviolet catastrophe, Heisenberg suggested that
avoiding local excitations of the vacuum (caused by interacting theories with
a maximal velocity) by sidestepping Lagrangian quantization and the ensuing
pointlike localized singular fields in favor of an S-matrix Ansatz could solve
the ultraviolet problem. His rather concrete proposals consisted in expressing
the unitary S-matrix in terms of a Hermitian phase operator $\eta.$ In modern
notation his proposal reads

$_{{}}$%
\begin{align}
S &  =\exp i\eta\\
\eta &  =%
{\displaystyle\sum}
\frac{1}{n!}%
{\displaystyle\int}
...%
{\displaystyle\int}
\eta(x_{1},...x_{n}):A_{in}(x_{1})...A_{in}(x_{n}):dx_{1}...dx_{n}\nonumber\\
&  \eta_{Hei}=g\int:A^{4}(x):d^{4}x\nonumber
\end{align}
where the on-shell coefficient functions of $\eta$ are chosen to be
Poincar\'{e} invariant and subject to further physically motivated
restrictions. In fact one such restriction which he suggested was that the
on-shell $\eta$ should be close to a Lagrangian interaction i.e. have local
coefficient functions as illustrated in the third line. It is customary to
split off the identity operator from $S$ and formulate unitarity in terms of a
quadratic relation for the T-operator
\begin{align}
S &  =1+iT\\
iT-iT^{\ast} &  =TT^{\ast}\nonumber
\end{align}
In this form the unitarity is close to the optical theorem and convenient for
perturbative checks.

Unitarity and Poincar\'{e} invariance are evidently satisfied if the (possibly
singular) functions $\eta(x_{1}...x_{n})$ are Poincar\'{e} invariant, but what
about macro-causality? For spacelike separation one must require the so called
\textit{cluster factorization property}. If there are n+m particles involved
in the scattering (the sum of incoming and outgoing particles) and one forms k
clusters (again containing in and out) and then separates these clusters by
large spacelike translations, the S-matrix must factorize into the product of
k smaller cluster S-matrices referring each describing the scattering
associated with a cluster. For the simplest case of two clusters
\begin{equation}
\lim_{a\rightarrow\infty}\left\langle g_{1}^{a},..,g_{m}\left\vert
S\right\vert f_{1}^{a},..,f_{n}\right\rangle =\left\langle g_{1},..\left\vert
S\right\vert f_{1},..\right\rangle \left\langle ..g_{m}\left\vert S\right\vert
..f_{n}\right\rangle
\end{equation}
where the first factor contains all the a-translated wave packets i.e. the
particles in the first cluster and the second factor contains the remaining
wave packets. \ In massive theories the cluster factorization is rapidly
attained. \ This asymptotic factorization property is usually written in
momentum space as
\begin{equation}
\left\langle q_{1},..q_{m}\left\vert S\right\vert p_{1},..p_{n}\right\rangle
=\delta-contribution+products~of~lower~delta\text{ }contributions
\end{equation}
i.e. the S-matrix contains besides the \textit{connected} contribution the
disconnected parts which consists of products of connected amplitudes
referring to processes with a lesser number of particles. The connected parts
have the correct smoothness properties as to make the formulas meaningful.

For timelike separated clusters the fall-off properties for large cluster
separations are much weaker. In fact there are inverse power law corrections
in the asymptotic timelike cluster distance. With the correct $i\varepsilon$
prescription (the same as Feynman's, but here only in the large timelike
asymptotic limit) they define what is referred to as \textit{causal}
re-scattering or the causal one-particle structure; The correct singularity
structure prevents the use of QT to construct time machines\footnote{A model
which was later shown\cite{Swieca} to lead to timelike precursors (as the
result of the presence of complex \ poles) was the Lee-Wick model.}.

For explanatory purpose of causal re-scattering imagine a kinematical
situation where the third particle enters the future cone of an interaction
region of particle 1 and 2 \ a long time after the 1-2 interaction happened,
and then scatters with the outgoing first particle leaving particle 2
undisturbed. \ In the limit of \textit{infinite timelike separation} the
connecting line of 1 and 3 i.e. which describes the trajectory after 1 leaves
the first process and moves to the scattering region with 3, the 1-3
intermediate propagator must coalesce with a causal propagator. i.e.
asymptotically this must be the Feynman propagator which is the only
one-particle propagator consistent with the causal structure of the re-scattering.

Whereas the cluster factorization of a Heisenberg Ansatz is obeyed by only
imposing the connectedness property on the $\eta$ coefficient functions, it is
not possible to satisfy the causal one-particle structure with a finite number
of terms in $\eta;$ in fact no pure S-matrix scheme has ever been devised
which secures the validity of the causal one-particle structure in the
presence of unitarity.

At this point the weakness of a pure S-matrix approach as advocated by
Heisenberg becomes exposed since there are certain properties which one can
formulate for the matrixelements of S but the non-linearity of the unitarity
requirements prevent their on-shell implementation "by hand". It is off-shell
QFT and its asymptotic timelike convergence, better known as scattering
theory, which saves us for spending the rest of our days with S-matrix
tinkering. The QFT correlation functions are the natural arena for
implementing causality properties; the observables are Hermitian and not
unitary and the building up of S-matrix unitarity is part of the asymptotic
convergence whose existence is guarantied by the properties of the correlations.

This problem of causal re-scattering in a Heisenberg S-matrix setting, and
more generally in any pure S-matrix formulation, was what finally convinced
Stueckelberg \cite{Stue} that a pure S-matrix approach is not feasible.

The S-matrix is without doubt the most important observable concept in
particle physics, but it should remain the "crown" of the theory and not its
foundation nor its principal computational tool. This was at least the gist of
Stueckelberg's critique on Heisenberg's program when he pointed out that to
reconcile macro-causality with unitarity "by hand" (i.e. without a an
off-shell setting which naturally unites these seemingly ill-fitting on-shell
concepts) one runs into insoluble problems.

Interestingly enough Stueckelberg then combined his idea of the causal one
particle structure with postulating pointlike interaction vertices and in this
way came to Feynman rules several years before Feynman. For showing that this
prescription leads to on-shell unitarity, he lacked the elegance of the
formalism of QFT in which the on-shell unitarity (and all the other properties
of S) is derived from simpler properties of correlation functions.

A systematic step for step derivation from a covariant Tomonaga setting of
QFT, including the Schwinger or Feynman formalism of renormalization, and with
particular care concerning the perturbative connection between QFT and the
S-matrix, was given by Dyson. It was also Dyson who raised the first doubts
about the convergence of the renormalized perturbative series.

The conceptually opaque status of perturbation theory lend importance to a
purely structural derivations of particles properties and scattering data
directly from the quantum field theoretic principles. Without having
mathematically controllable models at one's disposal, structural arguments
became increasingly important. Since despite all the difficulties to construct
interacting models there was no problem to define in mathematical clear terms
what are the requirements which are characteristic for QFT, the proof of the
existence of an S-matrix including all its properties (e.g. dispersion theory,
high energy bounds,..) from those well formulated requirements of QFT was a
quite pragmatic endeavour, even though it was often referred to as
"axiomatic". At least the original motivation for engaging in axiomatic QFT
was driven by the pragmatic desire to go beyond divergent perturbative series.
One of the high points of that research was the derivation of Kramers-Kronig
type of dispersion relation and their experimental verification which
strengthened the case for locality of interactions up to the present.

All this was achieved less than a decade after Stueckelberg's criticism of a
\textit{pure} S-matrix approach and the discovery of renormalized perturbation
theory by Tomonaga, Schwinger, Feynman and Dyson and forms the backbone of the
LSZ and Haag-Ruelle scattering theory.

As indicated above, the basic simplification of the problem of macro-causality
for the S-matrix consisted in the realization that its representation as the
large time scattering limit \textit{defuses the rather intractable nonlinear
problem} of implementing macro-causality in the presence of unitarity by
delegating it to simpler linear (off-shell) properties for correlation
functions. The problem with the nonlinear unitarity condition is in such a two
step process delegated to the linear problem of demonstrating the existence of
an isomorphism between two in and out free fields and the macro-causality of
the S-matrix follows from the on-shell preservation of certain properties
being related to off-shell micro-causality and spectral properties (energy
positivity). The connection between off-shell local quantum physics and the
on-shell S-matrix also shows the futility to invert this connection via
scattering theory by hand (this does not prevent string theorist to
contemplate such off-shell extrapolations). An S-matrix fulfilling the
crossing property can however be shown to admit only one inverse scattering
solution\footnote{Such inverse scattering problems show very clearly the
conceptual advantage of formulating QFT in terms of spacetime-indexed nets of
algebras rather than in terms of pointlike field coordinatizations of the
Lagrangian quantization. The crossing symmetric S-matrix is not capable to
highlight individual field coordinatizations, it only fixes the local net.} if
one assumes that the formfactors of the local quantum theory are also bound
together by crossing \cite{unique}. Unfortunately such uniqueness arguments
have, apart from the family of two-dimensional factorizing models, not led to
concrete constructions.

There is another particle physics setting for which a Poincar\'{e} invariant
unitary macro-causal S-matrix arises through scattering theory in the large
time asymptotic limit: \textit{Direct Particle Interaction} (DPI). It forgoes
micro-causality and fields and only retains Poincar\'{e} covariance and
macro-causality. It is certainly more phenomenological than QFT since it
contains interaction functions instead of coupling strength.

The reason why it is mentioned here (even though we are not advocating its use
outside medium energy pion-nucleon physics) is because \textit{its very
existence} not only \textit{removes some prejudices and incorrect folklore}
(including the belief that relativistic particle interactions are necessarily
QFTs or that a clustering S-matrix matrix can only arise from a QFT setting),
but it also indicates what has to be added/changed in order to arrive from
particle interactions to a full QFT setting. In other words it exposes some of
the nuts and bolts behind the field theoretic elegance.

Relativistic QM of particles is based on the Born-Newton-Wigner localization,
whereas the causal localization of QFT, which incorporated the finiteness of
the propagation speed, is related to the Poincar\'{e} representation theory
via modular theory (next section). The B-N-W localization of wave packets is
sufficient for recovering the forward lightcone restriction for 4-momenta
associated with events which are separated by large distances. Although this
suffices to obtain a Poincar\'{e} invariant macro-causal S-matrix, it prevents
the existence of local observables and vacuum polarization. For a presentation
of the differences and their profound consequences see \cite{interface}.

This DPI scheme introduces interactions between particles within a
multiparticle Wigner representation theoretical setting by generalizing the
Bakamjian-Thomas two-particle interacting Poincar\'{e} generators
\cite{Co-Po}. But whereas in the nonrelativistic QM\ the additivity of the
interaction potentials trivializes the problem of cluster factorization, there
is now no such easy connection between the modification of the n-particle
Poincar\'{e} generators and the nature of the interactions. Nevertheless, by
using the notion of \textit{scattering equivalences} one can arrive at a
cluster factorization formula for the interacting Poincar\'{e} generators and
the Moeller operators and hence also the S-matrix \cite{Co-Po}\cite{interface}%
. A scattering equivalence consists in a unitary transformation which changes
the representation of the Poincar\'{e} generators but maintains the S-matrix.
In the Coester-Polyzou DPI scheme the iteratively defined (according to
particle number n) interacting Poincar\'{e} generators lack the large distance
additivity associated with clustering and a scattering equivalence
transformation carried out for each n rectifies this situation.

It is interesting to note that the use of scattering equivalences is not the
only difference to nonrelativisyic QM; DPI theories also do not admit
combining the different n's into a second quantization Fock formalism. This
last property is not independent of the necessity to use scattering
equivalences in order to implement clustering.

One starts with a B-T two-particle interaction and compute the 2-particle
Moeller operator and the associated S-matrix as a large time limit of
propagation operators. As in the nonrelativistic case the two-particle cluster
property is satisfied for short range two particle interactions. For 3 and
more particles the construction of cluster factorizing Poincar\'{e} generators
and S-matrices require the iterative application of scattering equivalences.
The so constructed 3-particle S-matrix clusters with respect to the 2 particle
S-matrix in the previous step and it also contains a 3-particle connected part
which vanishes if any one of the particles is removed to spacelike infinity.
It is interesting that the Poincar\'{e} generators as well as the S-matrix
always contain a nontrivial connected part, i.e. in contrast to
nonrelativistic scattering the occurrence of direct higher particle induced
interactions cannot be prevented.

In the original formulation the scattering was purely elastic, but later it
was shown that an extension with particle creation channels is possible. Hence
the characteristic difference of DPI to QFT is not the presence of
creation/annihilation channels but rather the inexorable presence of
interaction-induced infinite vacuum polarization clouds in QFT.

As mentioned before such a scheme is purely phenomenological since the
interactions are not given in terms of coupling constants but rather coupling
functions \cite{Co-Po}.

An S-matrix with all the above properties fulfills the requirements of a
conjecture by Weinberg \cite{W}. But the S-matrix is not that of a QFT and
does not even agree for low energies with that coming from a QFT. Nuclear
physicists introduced this scheme precisely because quantum chromodynamics at
those energies does not permit any nonperturbative treatment and they wanted
to have a approximation scheme which is not completely phenomenological i.e.
at least not at variance with those macro-causality principles which can be
formulated for systems containing two nucleons in the presence of a small
number of mesons at relativistic energies (which allow already for meson
creation). In fact the DPI setting strictly is a S-matrix theory because
off-shell there are no covariant objects as conserved currents unless one
constructs them "by hand" (i.e. they are not natural objects within the DPI
formalism). Their introduction would require to go significantly beyond the
DPI scheme. The same applies to the incorporation of particle creation which
also has to be introduced by hand through the additional coupling of creation channels.

\textit{QFT and DPI are the only known settings for which an S-matrix with the
above properties can be derived} and which also have been reasonably well
understood from a conceptual/mathematical viewpoint. For DPI the mathematical
existence of models and their construction is handled in terms of well-known
functional analysis concepts as in ordinary QM. In case of QFT this is much
more difficult in view of the fact that the perturbative series is divergent
and the sometimes provable Borel resummability does not by its own establish
existence. Therefore it is deeply satisfying that its most intrinsic (field
coordinatization-independent) formulation in terms of spacetime localized
operator algebras has led to an nonperturbative existence proof and an
explicit construction of formfactors and the S-matrix for the special family
of interacting factorizable models. Hopefully this will be the beginning of a
new nonperturbative understanding which at the end may turn out to be the
realization of an intrinsic QFT without quantization "crutches".

The main aim of this article is to put forward arguments showing that string
theory is not what most people think it is and for what it received its name
namely a theory of an infinite collection of particles (a particle tower)
whose mass spectrum originates from a string which vibrates in spacetime. The
idea that it generalizes the pointlike localized fields of QFT is a metaphor
based on the mass spectrum which has no intrinsic meaning in the setting of
quantum localization. Since localization is a notorious difficult issue which
led to many misunderstandings, the discussion of localization of the objects
of string theory requires careful preparation which will be the main theme of
the next section.

Even in the simplest case of a free Nambu-Goto Lagrangian the facts contradict
the picture of dealing with stringy objects in spacetime: the string fields
associated with the \textit{N-G model is in fact pointlike localized}. Its
main difference to Lagrangian QFT is that the N-G string field\footnote{Sorry
reader, but one has to respect terminology, even if it was premature. One can
criticize a theory, but one cannot change its terminology.} behaves like a
generalized free field which has many more degrees of freedom than a standard
Lagrangian quantized field i.e. which accounts for the abundance of degrees of
freedom in an infinite particle tower. So the classical string aspect of the
N-G Lagrangian interpreted as a Lagrangian for a quantum field consists in
producing a generalized free field, apart from one tachyon component. For the
supersymmetric string which is free of tachyons one has to work with the
graded commutator.

The failure of implementing genuine string localization casts serious doubts
on the meaning of implementing interactions via the splitting and
recombination string associated tubes. The latter method avoids the coupling
of string fields (in analogy to the coupling of pointlike fields) which has
not led to useful insight. Instead one replaces it by something which is
formulated on the level "first quantization". An interaction for a finite
number of strings is then implemented "by hand" i.e. instead of doing this in
a Lagrangian setting, one extracts analytic expressions from Euclidean tube
pictures in analogy to the conversion of Feynman diagrams into the
perturbative analytic contributions of QFT. There is of course a significant
conceptual difference, whereas in the QFT setting these are recipes which can
be rigorously derived within a well-defined conceptual framework, string
theory leaves one empty handed, even after more than 4 decades of its existence.

In order to justify the tube picture as an analogy of particle physics string
theorists create a \textit{fake world of functional integral representations
of the quantum theory of the one-particle spaces}. Nobody had ever done
relativistic particle theory in such a obscure setting as that proposed by the
string theorists as a pedagogical warm up to the functional treatment of
string theory. It throws the cristal-clear and complete representation
theoretical classification of Wigner into the conceptual mud of extracting
infinite measure factors of the diffeomorphism group and similar ill-defined
manipulations. This is a significant step behind Wigner's representation
theoretical construction which is totally intrinsic and does not dependent on
artistry as extracting infinite gauge group factors not to mention the
completeness of Wigner's approach. This analogy does not reveal anything about
the locality of the "tubism" of string theory. Even if one closes all eyes
with respect to problems of localization, there remains the unsolved problem
whether this recipe defines at all a unitary macro-causal S-matrix because in
contrast to QFT there is no conceptual basis for why this should be expected.

Nevertheless the critique of string theory cannot be reduced to a simple
mistake which can be explained on one page, it rather needs $_{{}}$some
preparation on the conceptual as well as on the historical side. This is the
purpose of the next two sections.

\section{ On-shell crossing and thermal properties from causal localization}

In order to attain a solid vantage point for a critique of string theory, it
is necessary to recall the issue of localization which constitutes the basis
for the formulation and \textit{interpretation} of local quantum physics. The
easiest access with the least amount of pre-knowledge is through the Wigner
one-particle theory. .

Wigner discovered \cite{Wig} that irreducible positive energy ray
representations of the Poincar\'{e} group come in 3 families: massive
particles with half-integer spin, zero mass halfinteger helicity
representations and zero mass "infinite spin" representations. For brevity we
will refer to the families using numbers 1,2,3. Whereas the first and the
third family are rather large because their Casimir invariants have a
continuous range \footnote{Whereas for the massive family this is the value of
the mass operator, the continuous value in case of the infinite spin family is
the Casimir eigenvalue of the faithfully represented Euclidean group $E(2)$
(the little group of a lightlike vector).}, the finite helicity family has a
countable cardinality labeled by the halfinteger helicities. All up to present
in the laboratory observed particles are in the first two families. The fact
that no objects have been observed which fit into the third family should not
trick us into dismissing these positive energy representations since the
nature of observed dark matter is still unknown \cite{invisible}. Here the
third kind objects mainly serve the purpose to explain what
\textit{indecomposable string-like localization} means.

The three families have quite different causal localization properties. Let us
first look at the one with the best (sharpest) localization which is the
representation family of massive particles. For pedagogical simplicity let us
consider the Wigner representation of a scalar particle with the
representation space
\begin{align}
&  H_{Wig}=\left\{  \psi(p)|\int\left\vert \psi(p)\right\vert ^{2}\frac
{d^{3}p}{2p^{0}}<\infty\right\} \\
&  \left(  \mathfrak{u}_{Wig}(a,\Lambda)\psi\right)  (p)=e^{ipa}\psi
(\Lambda^{-1}p)\nonumber
\end{align}
We now define a subspace which, as we will see later on, consists of wave
function localized in a wedge. We take the standard $t-x$ wedge $W_{0}%
=(x>\left\vert t\right\vert ,~x,y$ arbitrary) and use the $t-x$ Lorentz boost
$\Lambda_{x-t}(\chi)\equiv\Lambda_{W_{0}}(\chi)$%
\begin{equation}
\Lambda_{W_{0}}(\chi):\left(
\begin{array}
[c]{c}%
t\\
z
\end{array}
\right)  \rightarrow\left(
\begin{array}
[c]{cc}%
\cosh\chi & -\sinh\chi\\
-\sinh\chi & \cosh\chi
\end{array}
\right)  \left(
\begin{array}
[c]{c}%
t\\
z
\end{array}
\right)
\end{equation}
which acts on $H_{Wig}$ as a unitary group of operators $\mathfrak{u}%
(\chi)\equiv$ $\mathfrak{u}(0,\Lambda_{z-t}(\chi))$ and the $x$-$t$ reflection
$j:$ ($x,t)\rightarrow(-x$,$-t)$ which, since it involves time reflection, is
implemented on Wigner wave functions by an anti-unitary operator
$\mathfrak{u}(j).$ One then forms the unbounded\footnote{The unboundedness of
the $\mathfrak{s}$ involution is of crucial importance in the encoding of
geometry into domain properties.} \textquotedblleft analytic
continuation\textquotedblright\ in the rapidity $U_{Wig}(\chi\rightarrow
-i\pi\chi)$ which leads to unbounded positive operators. Using a notation
which harmonizes with that of the modular theory in mathematics \cite{Su}, we
define the following operators in $H_{Wig}$
\begin{align}
&  \delta^{it}=U_{Wig}(\chi=-2\pi t)\equiv e^{-2\pi iK}\label{pol}\\
\mathfrak{\ }  &  \mathfrak{s}=\mathfrak{\ \mathfrak{j}}\delta^{\frac{1}{2}%
},\mathfrak{\mathfrak{j}}=U_{Wig}(j),~\delta=\delta^{it}|_{t=-i}\nonumber\\
&  ~\left(  \mathfrak{s}\psi\right)  (p)=\psi(-p)^{\ast}\nonumber
\end{align}
Since the anti-unitary operator $\mathfrak{j}$ is bounded, the domain of
$\mathfrak{s}$ consists of all vectors which are in the domain of
$\delta^{\frac{1}{2}}.$ With other words the domain is completely determined
in terms of Wigner representation theory of the connected part of the
Poincar\'{e} group. In order to highlight the relation between the geometry of
the Poincar\'{e} group and the causal notion of localization, it is helpful to
introduce the real subspace of $H_{Wig}$ (the closure refers to closure with
real scalar coefficients).%

\begin{align}
\mathfrak{K}  &  =\overline{\left\{  \psi|~\mathfrak{s}\psi=\psi\right\}
}\label{K}\\
dom\mathfrak{s~}\mathfrak{=K}  &  +i\mathfrak{K},~\overline{\mathfrak{K}%
+i\mathfrak{K}}=H_{Wig},\mathfrak{K}\cap i\mathfrak{K}=0\nonumber
\end{align}

The reader who is not familiar with modular theory should notice that these
modular concepts are somewhat unusual and very specific for the important
physical concept of causal localization; despite their physical significance
they have not entered the particle physics literature. One usually thinks that
an \textit{unbounded} anti-unitary involutive ($\mathfrak{s}^{2}=1$ on
$dom\mathfrak{s}$) operator which has two real eigenspace associated to the
eigenvalues $\pm1$ is an absurdity, but its ample existence is the essence of
causal localization in QFT.

The second line (\ref{K}) defines a property of an abstract real subspace
which is called \textit{standardness} and the existence of such a subspace is
synonymous with the existence of an abstract $\mathfrak{s~}$operator.

The important analytic characterization of modular wedge localization in the
sense of pertaining to the dense subspace $dom\mathfrak{s}$ is the strip
analyticity of the wave function in the momentum space rapidity $p=m(ch\chi
,p_{\perp},sh\chi).$ The requirement that such a wave function must be in the
domain of the positive operator $\delta^{\frac{1}{2}}$ is equivalent to its
analyticity in the strip $0<\chi<i\pi$ and the action of $\mathfrak{s}$
(\ref{pol}) relates the particle wave function on the lower boundary of the
strip which is associated to the antiparticle wave function on the negative
mass shell. This relation of particle to antiparticle wave functions is the
conceptual germ from which the most fundamental properties of QFT, as
crossing, existence of antiparticles, TCP theorem, spin-statistics connection
and the thermal manifestation of localization originate. Apart from special
cases this fully quantum localization concept cannot be reduced to supprt
properties of classical test functions.

More precisely the modular localization structure of the Wigner representation
theory "magically" preempts all these properties of a full QFT on the level of
the Wigner representation theory; to be more specific: these one-particle
properties imply the corresponding QFT properties via time-dependent
scattering theory \cite{Mu}. Hence any modification of any of those
fundamental properties (e.g. crossing---%
$>$%
Veneziano duality) is changing the principles of local quantum physics on
which are the result of more than half a century of successful particle
physics and therefore needs very strong scrutiny.

The mentioned one-particle indication of a thermal manifestation follows
directly from (\ref{pol}) by converting the dense set $dom\mathfrak{s}$ via
the graph norm of $\mathfrak{s}$ into an Hilbert space in its own right
$H_{G}\subset H_{Wig}$%
\begin{align}
\left\langle \psi\left\vert 1+\delta\right\vert \psi\right\rangle  &
=\left\langle \psi|\psi\right\rangle _{G}\label{an}\\
\left\langle \psi|\psi\right\rangle |_{dom\mathfrak{s}}  &  =\int\frac{d^{3}%
p}{2p_{0}}\frac{1}{1+e^{2\pi K}}\left\vert \psi_{G}(p)\right\vert ^{2}%
,~\psi_{G}\in H_{G}\nonumber
\end{align}
This formula represent the restriction of the norm to the strip analytic
function in terms of Hilbert space vectors $\psi_{G}$ which are free of
analytic restrictions. The result is the formula for a one point expectation
value in a thermal KMS state with respect to the Lorentz boost Hamiltonian K
at temperature 2$\pi$. As we will see in a moment, the modular relation
(\ref{pol})\ in the Wigner one-particle setting is the pre-stage for the
crossing relation as well as an associated KMS property in an interacting
QFT\footnote{The thermal manifestation of localization is the strongest
seperation between QM and QFT \cite{interface}.}.

Before we get to that point we first need to generalize the above derivation
to all positive energy representations and then explain how to get to the
sub-wedge modular localization for compact regions. For the generalization to
all positive energy representations we refer the reader to \cite{B-G-L}
\cite{MSY}, but since the sharpening of localization is very important for our
critique of string theory in the next section, it is helpful to be somewhat
explicit on this point.

In the first step one constructs the "net" of wedge-localized real subspaces
$\left\{  \mathfrak{K}_{W}\right\}  _{W\in\mathcal{W}}.$ This follows from
covariance applied to the reference space$\mathfrak{K}_{W_{0}}.$ In the second
step one aims at the definition of nets associated with tighter localization
regions via the formation of spatial intersections
\begin{equation}
\mathfrak{K}(\mathcal{O})\equiv\cap_{W\supset\mathcal{O}}\mathfrak{K}_{W}
\label{inter}%
\end{equation}
Note that the causally complete nature of the region is preserved under these
intersections in accordance with the causal propagation principle which
attributes physical significance to the causal closures of regions (this is
the reason for the appearance of noncompact or compact conic regions in local
quantum physics). In this way localization properties have been defined in an
intrinsic way i.e. separate from support properties of classical test functions.

The crucial question is how "tight" can one localize without running into the
triviality property $\mathfrak{K}(\mathcal{O})=0.$ The answer is quite
surprising: For all positive energy representations one can go down from
wedges to spacelike cones $\mathcal{O=C~}$of arbitrary narrow size
\begin{align}
&  \mathfrak{K}(\mathcal{C})~is\text{ }standard\\
&  \mathcal{C}=\left\{  x+\lambda\mathcal{D}\right\}  _{\lambda>0}\nonumber
\end{align}
i.e. the non-compact spacelike cones result by adding a family of compact
double cones with apex $x$ which arise from a spacelike double cone
$\mathcal{D}$ which touches the origin. Since there are three families of
positive energy Wigner representation\footnote{In d=1+2 there are also
plektonic/anyonic representations which will not be considered here.} one can
ask this question individually for each family.

The family with the most perfect localizability property is the massive one,
because in that case each $\mathfrak{K}(\mathcal{D})$ for arbitrary small
double cones is standard. On the opposite side is the third (massless infinite
spin) family for which the localization in arbitrarily thin spacelike cones
(in the limit semiinfinite strings) cannot be improved \cite{Yn}. The second
family (massless finite helicity) is in the middle in the sense that the
$\mathfrak{K}(\mathcal{D})$ spaces are standard but that the useful
"potentials" (vector potential in case of s=1) are only objects in Wigner
representation space if one permits spacelike cone localized objects i.e. they
covariant vectorpotentials cannot be associated with compact spacetime regions.

In fact there exists a completely intrinsic argument on the level of subspaces
associated with field strengths which attributes a representation theoretical
property to these "stringlike" potentials. It turns out that\ "duality"
relation (Haag duality)
\begin{equation}
\mathfrak{K}(\mathcal{O})=\mathfrak{K}(\mathcal{O}^{\prime})^{\prime}%
\end{equation}
in massive representations holds for all spacetime regions including
non-simply connected regions. Here the dash on $\mathcal{O}$ denotes the
causal disjoint, whereas $\mathfrak{K}(\mathcal{O})^{\prime}$ is the
symplectic complement of $\mathfrak{K}(\mathcal{O})$ in the sense of the
symplectic form defined by the imaginary part\footnote{For halfinteger spin
there is a slight change.} of the inner product in $H_{Wig}$ This ceases to be
the case for zero mass finite helicity representation where there is a duality
defect as soon as $\mathcal{O}$ is multiply connected (example: the causal
completion of the inside of a torus at t=0)$.$ In that case one finds
\begin{equation}
\mathfrak{K}(\mathcal{O})\subsetneqq\mathfrak{K}(\mathcal{O}^{\prime}%
)^{\prime}%
\end{equation}
which can be shown to be related to the string-like localization of potentials
\cite{MSY} i.e. this defect is the intrinsic indicator of the presence of
stringlike potentials.

These properties of localized Wigner subspaces can easily be converted to the
corresponding properties of a system (net) of spacetime indexed subalgebras of
the Weyl algebra or (for halfinteger spin) the CAR algebra. Since the reaction
between subspaces and subalgebras is functorial, all spatial properties have
their operator algebraic counterpart and one obtains (for simplicity we
restrict to the bosonic case)
\begin{align}
\mathcal{A(O})  &  \equiv alg\left\{  e^{i(a^{\ast}(\psi)+h.c.)}|~\psi
\in\mathfrak{K}(\mathcal{O})\subset H_{Wig}\right\} \label{mod}\\
SA\left\vert 0\right\rangle  &  =A^{\ast}\left\vert 0\right\rangle
,~A\in\mathcal{A(O}),~S=J\Delta^{\frac{1}{2}}\nonumber\\
\Delta^{it}\mathcal{A(O})\Delta^{-it}  &  =\mathcal{A(O}),~J\mathcal{A(O}%
)J=\mathcal{A(O})^{\prime}=\mathcal{A(O}^{\prime})\nonumber
\end{align}
It is important to not to misread the Weyl algebra generator in the first line
as an exponential of a smeared field; it is rather a (momentum space) Wigner
creation/annihilation operator integrated with Wigner wave functions from
$\mathfrak{K}(\mathcal{O})$ i.e. the functor uses directly the modular
localization in Wigner space and does not rely on smeared fields. The
antiunitary involution $J$ not only maps the algebra in its commutant (a
general property of the T-T modular theory) but, as a result of Haag duality,
also brings the causal commutativity into the game. Modular theory in the
general operator algebra setting leads to the modular group Ad$\Delta^{it}$
which leaves the algebra invariant and the antiunitary involution which
transforms the algebra into its Hilbert space commutant; both operators result
from the polar decomposition of the so-called (unbounded) Tomita involution
$S$. \ The field generators of this net of algebras are of course the
well-known singular covariant free fields whose systematic group theoretical
construction directly from the Wigner representation theory (except the
massless infinite spin representations) can be looked up in the first volume
of \cite{Wei}.

For a profound confrontation with string theory, the third Wigner
representation family is particularly useful. The history of its unravelling
is a very interesting illustration of the intricacies of localization
\cite{interface}, but in order not to loose time let us immediately pass to
the final result which consists in the realization that these third kind
Wigner representations cannot be point-localized. Unlike the finite helicity
representation these for which certain tensor fields (for s=1
vectorpotentials) are stringlike localized objects in an otherwise pointlike
generated representation, the wave functions of the third kind Wigner
representations are not compactly modular localizable i.e. all compact
intersections of wedge localized spaces are trivial and the smallest
noncompact intersections which still lead to standard $\mathfrak{K}(O)$ are
spacelike cones i.e. $\mathcal{O=C}$. The Weyl functor maps the spacelike
cone-localized subspaces directly into spacelike cone-localized operator algebras.

To make contact with the standard field formalism one looks at the
(necessarily singular) generators of these algebras. For the first two
families these are pointlike covariant fields $\Psi(x)$ apart from the finite
helicity potentials which, similar to the generators of the infinite spin
class, are described by string-localized field generators $\Psi(x,e)$ (leaving
off the tensor/spinorial indices) which depend in addition to a point x in
d-dimensional Minkowski spacetime also on a point in a d-1 dimensional de
Sitter space (the spacelike string direction) $e.$ The stringlike localization
nature shows up in the support properties of the commutator for whose
vanishing it is not sufficient that starting point x and x ' are spacelike but
rather
\begin{equation}
\left[  \Psi(x,e),\Psi(x^{\prime},e^{\prime})\right]  =0\text{~}%
only\text{~}for~x+\mathbb{R}_{+}e~><~x^{\prime}+\mathbb{R}e^{\prime}
\label{string}%
\end{equation}
The basic difference between the second (finite helicity) and third Wigner
representation type is that the string field generates subalgebras which are
generated by pointlike composites, whereas in case of the third (infinite
spin) type there is no pointlike localized composite. The theory also says
that there is no need to introduce generators which have a higher dimensional
localization beyond point- or semiinfinite string-like. Note that it is of
course not forbidden to introduce decomposable string (and higher) localized
operators as e.g.
\begin{equation}
\int\Psi(x)f(x)d^{4}x,\text{ }suppf\subset tube
\end{equation}
in the limit where the thickness of the tube approaches zero. When we talk
about semiinfinite string localization without further specification we mean
indecomposable strings. These are strings which in contrast to decomposable
strings cannot be observed in a counter since any registration device would
inevitably partition the string into the part inside and outside the counter
which contradicts its indecomposable nature (this is of course a metaphorical
argument which is in urgent need of a more explicit and intrinsic
presentation). The string-localized generators of the Wigner infinite spin
representation do not even admit pointlike localized composites i.e. net of
spacelike cone localized algebras has no compactly localized nontrivial
subalgebras. \ A milder form of string-like generation of representations
occurs for the zero mass finite helicity representation family which in some
way behaves localization-wise as standing in the middle between massive
representation (which are purely point-localized) and the third kind. These
representations are fully described in terms of pointlike localized field
strength but already before using these representations in interactions it
turns out that the additional introduction of "potentials" is helpful. Whereas
in the interaction free case there is a linear relation between the observable
field strength and its potential whose inversion permits to rewrite the latter
as one or more line integral over the former, this feature is lost under
suitable interactions i.e. the string localized potential may become an
indecomposable string localized generator which cannot be approximated by
compactly observables and therefore remains invisible to particle counters.

In the presence of interactions there is no \textit{direct} algebraic access
to problems of localization from the Wigner one-particle theory. In the
Wightman setting based on correlation functions of pointlike covariant fields,
the modular theory for the wedge region has been derived a long time ago by
Bisognano and Wichmann and more recently within the more general algebraic
setting by Mund\footnote{That derivation actually uses the modular properties
of the Wigner setting which is connected via scattering theory to the
interacting wedge-localized algebras and then as explained above (via
intersection) to the modular structure of all local algebras $\mathcal{A(O}%
).$} \cite{Mu}. The resulting modular S-operator has the same property as in
(\ref{mod}) i.e. the "radial " part of the polar decomposition of the modular
involution $S$ is determined solely by the representation theory of the
Poincar\'{e} group i.e. the particle content whereas the $J$ turns out to
depend on the interaction \cite{Ann} since it is related to the scattering
operator $S_{scat}$%
\begin{equation}
J=J_{0}S_{scat} \label{J}%
\end{equation}
which in this way becomes a relative modular invariant between the interacting
and the free wedge algebra\footnote{$J_{0}$ is (apart from a $\pi$-rotation
around the z-axis of the t-z wedge) the TCP operator of a free theory and $J$
is the same object in the presence of an interaction.}. There is no change in
the construction of the $\mathcal{A(O})$ by intersecting $\mathcal{A(}W)s.$
However in the presence of interactions the functorial relation between the
Wigner theory gets lost. In fact no subwedge-localized algebra contains any
associated PFG (polarization-free-generator) i.e. an operator which creates a
one particle state from the vacuum without an additional vacuum polarization
cloud consisting of infinitely many particle-antiparticle pairs.

Since the crossing property played a crucial role in S-matrix approaches to
particle physics, it pays to spend some time for its appropriate formulation
and on its conceptual content. Its most general formulation is given in terms
of formfactors which are products of W-localized operators $A_{i}%
\in\mathcal{A}(W)$\footnote{Since all compactly localized operators can be
translated into a common W and since the spacetime translation acts on in and
out states in a completel known way this is hardly any genuine restriction.}
between incoming ket and outgoing bra states
\begin{align}
&  ^{out}\left\langle p_{k+1},p_{k+2},...p_{n}\left\vert A\right\vert
p_{1},p_{2},..p_{k}\right\rangle ^{in}=\label{cross}\\
&  ^{out}\left\langle -\bar{p}_{k},p_{k+1},p_{k+2},...p_{n}\left\vert
A\right\vert p_{1},p_{2},..p_{k-1}\right\rangle ^{in},~A=\Pi_{l}A_{l}\nonumber
\end{align}
where the crossed particle is an outgoing anti particle relative to the
original incoming particle Hence all formfactors of $A$ with the same total
particle number n are related to one \ "masterfunction" by analytic
continuation through the complex mass shell from the physical forward shell to
the unphysical backward part. Hence the predictive power of crossing is
inexorably connected with the concept of analytic continuation i.e. it is
primarily of a structural-conceptual kind. It is convenient to take as the
master reference formfactor the vacuum polarization components of $A\Omega$
i.e. the infinite system of components of the infinite vacuum polarization
cloud of $A\Omega.$ Needless to add that the crossing relation may be empty in
case that the operator $A~$cannot absorb the energy momentum difference
between the original value and its continued negative backward mass shell
value. In this setting the S-matrix arises as a special case for
$A=\mathbf{1}$ i.e. an operator which cannot absorb any energy momentum. In
this case it is not possible to use the vacuum polarization as a reference and
neither leads the crossing of one momentum in the 2-particle elastic amplitude
to a meaningful relation (but the simultaneous crossing of two particles in
the in and out configuration is meaningful).

This is also the right place to correct the cartoon picture of the QFT vacuum
as a bubbling soup which for short times, thanks to the Heisenberg uncertainty
relation between time and energy, can violate the energy momentum
conservation\footnote{The origin of these metaphors sees to be the too literal
interpretation of the momentum space Feynman rules.}. This is metaphorical
humbug; the correct picture is that (modular) localization in QFT costs
energy-momentum i.e. in order to split the vacuum into a tensor product with
controllable vacuum fluctuations
\begin{align}
\Omega &  \rightarrow\Omega_{1}\otimes\Omega_{2}\\
A  &  \rightarrow A_{1}\otimes A_{2}\nonumber
\end{align}
where the index 1 refers say to a compact region e.g. a double cone and 2
refers to is noncompact causal complement one must one must spend an unbounded
amount of energy so that the vacuum after the split becomes thermal whereas in
QM (the home of the uncertainty relation) the tensor-product splitting of a
global system into a box and its complement does not cost anything because
localizing in the sense of Born is for free; it is a mental process which is
related to information and has nothing to do with thermal properties. The
conceptual difference between Born- and modular- localization is considerable
and the bad habit of confusing the two is the main cause for the "information
paradox". \ 

It is not only string theory which dwells on metaphors, but some of those in
QFT e.g. the above mentioned bubbling vacuum at least do not course any
serious damage. Another is the idea with strong connection to metaphors is
that physical energy momentum is simply the Fourier transform of an x which
behaves covariant under translations. The physically correct definition is
through the geometric relation between counter events separated by an
asymptotically large distance as in scattering theory. The realization that
Feynman rules become useless in perturbative situations without spacetime
symmetries as for QFT in curved space and that case the perturbative
expressions must obey the subtle property of the recently formulated local
covariance principle took more than 3 decades to become appreciated.

The origin of the formfactor crossing property lies in the strip analyticity
of wedge localized states and correlation function. For wedge localized wave
functions this was explained above (\ref{pol}, \ref{an}). For simplicity let
us limit the interacting situation to the simplest case
\begin{align}
\left\langle 0\left\vert A\right\vert p\right\rangle  &  =\left\langle
-\bar{p}\left\vert A\right\vert 0\right\rangle \\
\left\langle 0\left\vert AB\right\vert 0\right\rangle  &  =\left\langle
0\left\vert B\Delta A\right\vert 0\right\rangle \nonumber
\end{align}
where in the second line we have written the KMS property for the wedge
algebra which is a general consequence of modular operator theory and for the
special case of wedge localization agrees with Unruh's observations about
thermal aspects of Rindler localization ($\Delta^{it}=U_{W}^{boost}(\chi=-2\pi
t)$). But how to view the first relation as a consequence of the second? The
secrete is that although the intersection of the space of one-particle states
with that obtained from applying compact localized algebras to the vacuum (and
closing in the modular graph norm) is trivial, that with the noncompact
wedge-localized algebra is not; it is even dense in the Hilbert space. Once it
is understood that there exists a wedge affiliated operator $B$ which, if
applied to the vacuum, generates the one-particle state, one can apply the KMS
relation in the second line. The rest the follows from transporting the left
side $B$ as $B^{\ast}$ to the bra vacuum. The rest follows by rewriting the
$B^{\ast}\left\vert 0\right\rangle $ as $SB\left\vert 0\right\rangle $ using
modular operator theory and using (\ref{J}). The resulting $\Delta^{\frac
{1}{2}}JB\left\vert 0\right\rangle =\Delta^{\frac{1}{2}}J_{0}B\left\vert
0\right\rangle $ (since the $S_{scat}$ matrix acts trivially on one-particle
states) leads to the desired result\footnote{The plane wave relation should be
understood in the sence of wave packets from the dense set of strip-analytic
wave functions.} $\Delta^{\frac{1}{2}}J_{0}\left\vert p\right\rangle
=\left\vert -p\right\rangle .$ The general form (\ref{cross}) would follow if
we could generalize the KMS relation to include operators from the wedge
localized in and out free field algebras. Although they share with
$\mathcal{A}(W)$ the same unitary Lorentz boost as the modular group, their
modular inversions $J$ are not equal and hence additional arguments are
required. We will leave the completion of the derivation of crossing to a
future publication \cite{M-S}.

Fortunately in order to criticize the string theory interpretation of the
canonically quantized Nambu-Goto mode we do not have to go into subtle
details. For such bilinear Lagrangians (leading to linear Euler-Lagrange
equations) the connection between localization of states and locality of
operators is that in free field theory. In this case it is possible to pass
from the "first quantized" version directly to its "second quantization" i.e.
to the N-G "string field theory". Since the physical content consists of an
infinite tower of massive particles (with one layer of finite helicity
massless representations), the only question is does the original classical
parametrization lead to fields which are decomposable strings or are they
point localized? In the first case one could resolve the composite string in
terms of a stringy spread of underlying pointlike fields (i.e. the Lagrangian
does not directly lead to pointlike objects) whereas in the second case the
terminology would not even have a metaphorical meaning. The suspense will be
left to the next section.

\section{A turn with grave consequences}

Although the protagonists of the S-matrix bootstrap placed the new and
important crossing property into the center of their S-matrix program, they
failed to come up with a constructive proposal which could implement this new
requirement. Other older requirements, as Stueckelberg's macro causality, was
not even mentioned in their program, they where probably forgotten in the
maelstrom of time. The important question in what way (on-shell) crossing is
related to the causality principles of QFT went against their ideology which
(at least in the later stages) was to cleanse particle physics from the
dominance of QFT. In fact most of their efforts were focussed on the elastic
scattering amplitude on which Mandelstam's conjecture concerning the validity
of a certain double spectrum representation was tested in terms of which the
crossing had a simple formulation.

The crossing property in the form of the previous section was first noticed in
the Feynman perturbation theory where a certain analytic continuation from
momenta on the forward to the backward shell only changes the in/out
association of the external legs of Feynman graphs within the same
perturbative order. Hence crossing relates Feynman graphs in a fixed order
with the same total number of legs where the distribution of the momenta
between forward/backward mass shell is changed by analytic continuation on the
complex mass shell. Whereas unitarity and macro-causality are relativistic
particle properties (in the sense that they can be formulated and implemented
without mentioning fields), the crossing property is a statement which uses
analyticity properties whose origin, as most analytic properties in particle
physics, is in local quantum physics and not in (relativistic) QM. Properties
of local quantum physics do not permit a natural description in terms of
operator tools of QM. The method which is intrinsic to their conceptual
structure is that of spacetime indexed operator algebras.

From looking at the original papers It is quite evident that Veneziano
\cite{Venez} had this kind of crossing in mind when he set out to construct an
explicit implementation within the Mandelstam setting for 2-2 elastic
scattering amplitudes. But being guided by the properties of $\Gamma
$-functions and the idea to implement crossing with one-particle poles alone
(supported by the pole-dominance of Regge pole phenomenology of the day) he
arrived at "Veneziano duality" which is different from the kind of crossing
which is an on-shell imprint of the causal locality principle. The S-matrix
(and more general the formfactor) crossing from QFT is a delicate interplay
between a finite number of one-particle poles and the scattering continuum.
The Veneziano duality on the other hand required the presence of a tower of
particles and no participation of the scattering continuum. The idea of
one-particle saturation of scattering amplitudes by an infinite particle tower
came from the Regge-pole phenomenology; Veneziano's merit was to have
recognized that this saturation idea harmonized perfectly with properties of
gamma functions. But unfortunately it is not compatible with QFT. The
consistency of crossing in QFT can be explicitly verified in factorizing
models \cite{Ba-Ka}. For the Veneziano duality one could point to string
theory neither its conceptual structure nor its mathematical status is known. functions.

Extending the search for an implementation of duality-based on properties of
Gamma function, Virasoro \cite{Vir} arrived at a model with a different and
somewhat more realistic looking particle content. The duality setting became
more completed and acquired some mathematical charm after in \cite{DHS} it was
extended to n particles. The resulting "dual resonance model" was the missing
link from the phenomenological use of Gamma function properties to a
conceptually and mathematically more attractive formulation in terms of known
concepts in chiral conformal QFT, the new idea being that Minkowski spacetime
should be envisaged as the "target space"\footnote{Note that the notion of
target space is well defined only in classical field theories (where fields
have numerical values) whereas in QFT its meaning is metaphorical.} of a
suitably defined chiral model.

It is worthwhile to look at the mathematical formulation and the associated
concepts in some detail. The conformal model which fits the dual resonance
model are the charge creating fields of a multi-component abelian chiral
current which are customarily described in the setting of bosonization
\begin{equation}
\Psi(z,p)=:e^{ip\phi(z)}:
\end{equation}
where the d-component $\phi(z)$ is the potential of a d-component chiral
current $j(z)=\frac{d}{dz}\phi(z),$ and p is a d-component numerical vector
whose components describe (up to a shared factor) the value of the charge
which the $\Psi$ transfers$.$ The fact that the Hilbert space for $\phi$ and
$\Psi$ is larger than that of the current $j$ is the only place where the
language of bosonization becomes somewhat metaphoric; this point is taken care
of by a proper quantum mechanical treatment of zero modes which appear in the
Fourier decomposition of $\phi(z).$ The n-point functions of the $\Psi$ are
the integrands of the scattering amplitudes; the latter result from the former
by z-integration after multiplication with $z_{i}$ dependent factor
\cite{Vec}. \ Hence it is more or less obvious that the energy-momentum
conservation of the target theory results from the charge conservation of the
conformal source theory. Even for the cluster property one finds a convincing
argument in that it corresponds to the cluster decomposition of the auxiliary
chiral theory (if one rewrites it in terms of the noncompact parametrization).
But it is too early to rejoice since there are still three hard problems
ahead: the macro-causal rescattering structure, unitarization and the origin
of the restriction to 10 charges for accommodating a Poincar\'{e} symmetry in
the target space.

Regarding the timelike macro-causality the problem is to avoid falling back
behind Stueckelberg. As far as unitarization goes the problem is to find a
structural argument which replaces the unitarity-securing scattering theory in
favour of some property of the auxiliary chiral conformal QFT.

Whereas the first dual model papers are hard to criticize since they represent
at best bits of a new S-matrix theory, the dual resonance model, although
being obviously not yet an S-matrix theory as it lacks unitarity, is already a
concrete target for criticism. According to is protagonists its n-particle
amplitudes should be interpreted as the tree approximation of an unknown
unitary S-matrix. The position of the one-particle poles and their residua in
the various channels obtained by crossing incoming and outgoing lines and vice
versa were shown to be consistent with the required duality property. This is
not only a nice mathematical achievement but, ignoring for a moment the
unitarity problem, it also secures the validity of the causal one particle structure.

Different from the representation theoretical approach to the Poincar\'{e}
symmetry a la Wigner, on which QFT is founded, the dual model realizes
Poincar\'{e} symmetry as \textit{a noncompact inner symmetry of a chiral
conformal QFT}. To be more precise the spacetime symmetry acts on the
field-value (target) space which in a Lagrangian quantization approach is the
arena for compact internal symmetry actions. The problem at hand is to
describe the dual model S-matrix in terms of an auxiliary chiral conformal so
that the momenta are the continuous values of an abelian multicomponent
current and that the Poincar\'{e} group acts unitarily on that spectrum of
multi-component charges. The first part of this requirement is fulfilled
thanks to the fact that the chiral abelian current theory has in contrast to
so called rational chiral models a continuous supply of superselection
sectors\footnote{On higher dimensions observable the representation theory of
local of local observables leads to charge sectors which can only accomodate
compact inner symmetry groups \cite{Ha}.} and so a match with a continuous
energy-momentum spectrum is possible. The second part of this requirement is
more difficult to enforce and indeed it is in general not possible to
accommodate a noncompact internal symmetry group which acts on the target
space of a QFT. The fact that it is possible to obtain a positive energy
representation on a suitably restricted target space in 10 spacetime dimension
if one extends the abelian charge chiral theory by supersymmetry has been
verified by computations. Another important property is the pole structure of
the dual model scattering amplitudes (the particle tower) which must originate
from some property of the conformal correlations in terms of which these
amplitudes are defined. Here one looks in vain for an explanation in terms of
the intrinsic logic of chiral (source!) model. But a clarification of this
point would be important if, as string theorists do, one wants to attach a
spacetime interpretation in the sense of a worldsheet carved out by a string.
This has not been done despite the fact that this picture is always inferred
and here begins the metaphoric twighlight.

\ Since the dual resonance model is the point of departure of string theory
and extra dimension there is the danger that a large part of physics of the
last two decades is based on metaphoric ideas with doubtful reality content.
This is why I believe that the lack of clarification of duality versus
crossing is one of the most important missed opportunities in the history of
physics. Whereas the field theoretic crossing cannot only be seen in
perturbation theory but also enters as an important tool in the
nonperturbative construction of factorising models, there is not a single
example for a dual theory (the dual resonance model is not a model for an S-matrix).

An equivalent formulation which has the advantage of permitting a
interaction-free Lagrangian presentation for the particle tower spectrum (with
the interactions added "by hand") was proposed by Nambu \cite{N} and Goto
\cite{G} and in a more standard functional integral setting by Polyakov
\cite{Pol}. This string theory setting confirmed the restriction of target
space to a 10-dimensional super Poincar\'{e} group. But is such an argument
acceptable as a prediction that we are living in a 10 dimensional space with 5
spatial dimensions curled up in such a way that they have escaped observation?
Can the answer to such a fundamental almost metaphysical question about the
dimensionality of our world be left to mind games which are orthogonal on
those symmetry principles which incorporate all our past observations? In
short, is string theory a metaphorical aberration or is it a gift of the 21
century which fell by luck into the 20 century? This is the main question
which will be pursued here. We will end this section by posing a series of
critical questions whose further pursuit will constitute an important part of
this article.

\begin{itemize}
\item In relativistic theories (whose existence is assumed) in which the
S-matrix arises as a long time limit there is no problem with unitarity and
macro-causality of $S.$ But what if the S-matrix, or rather what one thinks
should be its tree approximation, is the result of phenomenological tinkering
which has nothing to do with a large time limit in an underlying QFT? Is the
mathematical respectability of a mind game like this a sufficient strong
prerequisite for taking it serious as a construction principle in particle physics?

\item What is the physical meaning of a purely auxiliary concept as "string"
which enters the prescription for the construction of an S matrix. Is the
result of the quantization of a classical Nambu-Goto string really a string in
the sense of quantum localization; or more pointedly: does the well-known
parallelism between pointlike classical fields and their quantized
counterparts carry over to strings?

\item Does the presence of a zero mass spin=2 particle in a covariant theory
with only one parameter (the claimed finiteness of string theory) make it
automatically a candidate for QG? What about the local covariance principle
which secures background independence?
\end{itemize}

In the beginning of the 70s better experimental results on high transverse
momentum transfer scattering pulled the rug out from under the dual model and
all ideas related to Regge phenomenology and made it a conceptual orphan.
However some people who invested a lot of time and also some computational
ingenuity which led to these surprising (in the sense of not expected by the
intrinsic logic of abelian current models) results. The phenomenological start
took a sudden theoretical turn without being able to find a conceptual anchor.

\section{The ascend of the metaphoric approach to particle physics: string
theory}

During the first years of its existence the dual model and its various
extensions attracted some attention from the Regge phenomenology community; in
fact most of the dual model protagonists came from that area. With the ascend
of exciting new ideas about strong interactions coming from QCD, which also
offered a vast new playground for phenomenologists, the Regge pole era came to
an end and with new hadronic large-momentum transfer data arriving, which
contradicted the dual model predictions, the dual model formalism finally lost
its observational support. In this way it became an "orphan" of particle
physics since its modest mathematical charm, which consisted in Euler type of
identities between gamma functions, lost its physical
attraction\footnote{Interestigly enough representations of scattering
amplitudes in terms of Gamma functions re-appeared a decade later in
connection with two-dimensional factorizing models but this time in complete
harmony with unitarity and with the true (QFT) crossing (i.e. these models
have a finite number of particles and hence there is no duality).}.

Although the small community of dual modelists were unshaken in their belief
that hidden behind the many unexpected properties there exists a deep new kind
of quantum physics, they hardly made any effort to improve the conceptual
understanding of crossing and its relation to duality. This would have been
the right time to come to terms with the physical spacetime origin of the
on-shell crossing property and possible modifications needed to understand
Veneziano duality.

As explained in a previous section the standard crossing property is an
interplay between (a finite number of) particle poles and multiparticle cuts
in the analytic scattering amplitude. In particular there is no understanding
whether there is any physical property behind the formal manipulation of
satisfying the crossing by simply trading the continuum contribution with a
tower of particle poles. Veneziano's construction was not dictated by physical
necessity but rather by computational expediency:in the typical physicists way
according to the motto: if you cannot solve a problem coming from physical
principles, try to invent another similar looking one which you can solve and
try to interprete the outcome. It was obvious that the duality requirement
does not have a solution with a finite number of particles and it took the
mathematical ingenuity of Veneziano to construct a solution with infinitely
many particles. In view of the precarious conceptual status of this result
makes it understandable that the string interpretation was welcomed as a
conceptual salvation since it brought the speculative new ideas nearer to the
Lagrangian shore.

In view of the fact that duality was not a property of S-matrices of QFT, and
not a consequence of any known physical principle, but rather a result of
gamma function "tinkering" combined with "Reggeology", a profound conceptual
study of these observations was warranted; but the problem was not even posed.
In d=1+1 the nontrivial family of factorizing models provides explicit and
rigorous illustrations of crossing in QFT coming about as a subtle interplay
between one-particle poles and the scattering continuum \cite{Ba-Ka}%
\cite{Lech}. A factorizing model which fulfills duality based on infinitely
many one-particle poles does not exist.

In fact it was an uneasy feeling that the dual model was constructed with an
excess of sophisticated "tinkering" and a lack of guiding principles. This
theoretical frailty became more apparent, especially after the model lost the
protection which it enjoyed under the phenomenological Regge pole umbrella,
where the demands on conceptual coherence were less restrictive. Hence the
observation that those operator formulas representing the on-shell scattering
states of the dual resonance model could be viewed as coming from canonical
quantization of a classical relativistic string was considered as an act of
conceptual liberation which incorporated the dual model into an apparently
conceptually more satisfactory setting. With the picture of a relativistic
string in mind, there was hope to obtain an intrinsic access to interactions
and to complete the construction of a unitary S-matrix in such a way that the
dual resonance model is the lowest order in a new perturbative systematics.

This hope changed to ecstasy when some researchers became aware that in
contrast to higher spin ($s>1$) QFT which, even within a short distance
improving BRST ghost formalism, apparently leads to infinitely many
perturbatively undetermined parameters (nonrenormalizability), string theory
has basically only one parameter (assuming that it stays finite in every order
which nobody has been able to check).

Naturally the case of s=2 contribution attracted most attention as a result of
its promise to lead to a finite theory of quantum gravity. In contrast to QFT,
which in addition to perturbative series representation has a wealth of
strutural theorems, string theory offers nothing like this; perturbative
results are only secured (with tremendous computational effort and no gain in
physical insight) up to second order and nonperturbative statements (e.g.
statements about branes) are not available in the form of structural theorems
but remain in the realm of quasiclassical calculations and metaphoric
reasoning (even after more than four decades!). So the hope that (super)string
theory was the liberating act by which the old phenomenological ideas could be
elevated to a new TOE with QG as its shiny byproduct remained unfulfilled but
this did not prevent the formation of a large community around string theory.

Instead of entering a point for point critique of the extensive and
technically laborious content of string theory, I prefer to focus my critical
remarks to what I consider the Achilles heel of string theory, namely its
metaphoric relation with those localization concepts which are central for the
formulation and interpretation of particle physics.

We know from Wigner's representation theoretical classification that the
indecomposable constituents of positive energy matter are coming in 3
families: the massive family which is labeled by a continuous mass parameter
and a discrete spin, a discrete massless family with discrete helicity and
finally a continuous zero mass family of with an infinite spin (helicity)
tower. Whereas theories involving the first two families have generating
pointlike localized field strengths, there are no pointlike covariant
generators within the last family; rather the sharpest localized generators in
that case are semiinfinite strings localized along the spacelike half-line
$x+\mathbb{R}_{+}e,$ where $x$ is the starting point of the string and $e$ is
the spacelike direction in which it extends to spacelike infinity. Their
localization shows up in their commutation relation which we presented in
(\ref{string}).

Stringlike localized objects can of course also be constructed in pointlike
QFTs; one only has to spread a pointlike field along a string where the
spreading has to be done in the sense of distribution theory since the
resulting stringlike objects is also singular. Such a string will be referred
to as composite or \textit{decomposable}. It has no fundamental significance
and one would not expect such onjects from a Lagrangian setting, although its
use in special cases may be helpful in exploring the physical content of a
model. One very good technical reason for doing this is the fact that the
\textit{short distance behavior of a spacelike semiinfinite string localized
covariant field is better that of its pointlike counterpart}. A pointlike free
massive vector field $A$ has short distance scaling dimension $sddA=2$ whereas
its string localized counterpart has the lower value $sdd\Phi_{A}=1,$ in fact
the short distance dimension of stringlike massive fields of arbitrary high
spin remain at $sdd\Phi=1$ \cite{MSY}$.$

Hence a perturbation theory in terms of interactions between such decomposable
covariant string localized fields can be expected to enlarge the possibilities
for renormalizable interactions, in particular interactions involving higher
spin ($s\geq1$) for which the power counting test\footnote{Positivity (no
ghosts) is always assumed.} for renormalizability leaves no interactions at
all. Of course the inductive perturbation rules (the Epstein-Glaser rules) are
different from the pointlike case. The use of stringlike covariant fields is
still pretty much in its infancy. \ 

What matters here is that the representations of the third kind of positiv
energy Wigner matter are \textit{indecomposable strings}; in fact it is not
difficult to argue that the associated QFT has \textit{no pointlike localized
composite fields} and hence no compact localized operators at all
\cite{invisible}. These representations are therefore excellent illustrations
for the meaning of \textit{indecomposable stringlike localized fields.}

From our previous discussions we know that \textit{localization is an
autonomous quantum theoretical concept} which is governed by the
representation theory of the Poincar\'{e} group and not by trying to transfer
classical localization via quantization into the quantum realm.

It is true that \textit{both localization concepts coalesce in the pointlike
case}. This fact was extremely beneficial for the early birth of QFT more than
one decade before Wigner's path-breaking work on the intrinsic representation
theoretic method to classify particles as indecomposable objects. The
classical-quantum quantization parallelism was also crucial for the
development of the Lagrangian and functional approach. For pointlike
generators the quantization approach and Wigner's representation theoretical
method are largely equivalent. The most detailed account of this equivalence
can be found in Weinberg's book \cite{Wei} where the group theoretic
formulation of Poincar\'{e} covariance is used to construct a (countably
infinite) family of intertwiners which map the canonical Wigner representation
into the non unique covariant (undotted/dotted spinorial) representations.

The third kind of positive energy Wigner representation shows the limitation
of pointlike localization and Lagrangian quantization. It is evident that any
critique of string theory has to start with a profound review of its
localization. The main topic for the rest of this section will be to
demonstrate the correctness of the following theses: \textit{the}
\textit{objects} \textit{of}\ \textit{string theory are not string-localized
in any intrinsic quantum-physical sense}

The above representation theoretical discussion shows that massive states in
interaction-free Poincar\'{e}-invariant theories can only be string-localized
in a decomposable sense i.e. as a pointlike \ localized state spread over an
infinitely thin tube. But there is no possibility of having a massive
\textit{indecomposable} state which would be string-localized whereas a
decomposable string state can be written in terms of spread pointlike states.

We all have been exposed to the story of the "little wiggling strings"
(meanwhile there are also the large cosmic strings) in spacetime; this has
been the opening mantra with which string theorists usually introduce their
talks. Of course the reality content of metaphors is not enhanced by the fact
that the storytellers seriously believe in what they are saying. For free
strings, more precisely for the string field theory associated with the
Nambu-Goto string, this issue can and has been completely settled by an
explicit calculation of the commutator function which for all bilinear free
systems is a c-number which carries the full intrinsic information about
quantum localization.

This was done, first by string theorists \cite{Mar} and afterwards by
mathematical physicists \cite{Dim} and the result in both cases was a point-
and not a string-like localization. In other words there was neither a
composite nor a indecomposable string, rather it was pointlike\footnote{It
could have been that the classical string parametrization of the N-G
Lagrangian under canonical quantization passes to a decomposable (composite)
string in the aforementioned sense, but this is not the case.}. If not in the
commutator function then where else should the intrinsic meaning of a
spacetime string show up? This is precisely the point in the story from the
dual model to string theory where metaphors have won over the observable
content. According to string theorists the strings themselves remain
invisible, the commutator shows only their center of mass position. What is
even more surprising is that that we (quantum field theorists, mathematical
physicists) i.e. a mathematical physicist as \cite{Dim} who has carefully
checked the commutator and also finds a pointlike localization but also
concludes that for some reason the string itself remained invisible. This
shows \textit{the incredible spell of metaphors once they become accepted by a
sufficiently large community}. In such a sociological environment our often
praised scientific objectivity and independence suffers a meltdown (just like
public opinion in democratic states on carefully staged media-supported
bellicose policy of a governmental elite) and gives way to a preemted obedience.

As expected on the basis of covariance, the putative string field
\footnote{Like a monument becomes protected as part of history even if that
part of history was anything to be proud about, it is not possible to change
terminology in physics just because a misleading name was selected
prematurely. This requires the reader to pay attention to not confuse objects
of string field theory with string localised objects.} is really an pointlike
localized generalized free field with an infinite mass- and spin- spectrum
(i.e. a mass-spin tower). The pure bosonic N-G Lagrangian suffers from the
tachyon "flaw" i.e. the violation of the positive energy requirement; this is
removed by\ taking the supersymmetric extension of the N-G model which uses
the graded commutators.

The infinite mass tower together with the c-number nature of the graded
commutator resembles the spectrum of infinite component fields. The metaphoric
idea that this N-G model mass-spin spectrum which is reminiscend of a
classical string can be interpreted as a result of a string which vibrates in
spacetime is contradicted by direct operator calculations. The N-G string is
not even a decomposable string in the sense of a pointlike field spread over a
infinitely thin tube. Indecomposable stringlike localized objects occur in
Wigner's representation theory but they have nothing in $_{{}}$common with the
pointlike localized objects of string theory presented in (\ref{string}).

There is also another reason why in the face of all evidence to the contrary
string theorists cling to their string metaphors. It is precisely that
metaphoric language which helps them to define their interaction in terms of
graphical splitting and recombining euclidean tubes (string \ "tubism"). This
is to be expected since genuine (i.e. non transient) metaphors can only be
sustained at the cost of constructing more metaphors ("excuses" in the words
of Feynman, who in my view saw some of these problems coming). As long as one
accepts the first metaphor, there is no problem with interpreting the
interaction but without it there is no motivation.

\textbf{With the insistence of interpreting the result of the canonical
quantization of the Nambu-Goto Lagrangian as an (invisible, apart from the
center of mass) string-localized object in spacetime, the metaphoric rubicon
has been crossed. As a consequence the metaphoric style has become accepted
even on subjects of particle theory which are not directly related to string
theory. It is truly amazing to see that almost 90 years after quantum
mechanics begun with Heisenberg's placing observables into the center of the
new quantum reality in order to avoid the contradictions caused by the
contamination of the new quantum physics with classical pictures, the
metaphoric world view is back in strength and it seems to be even embraced by
the Zeitgeist.}

Metaphors in particle physics are useful as long as they remain transitory
devices in the sense of exploring new physics in a still poorly understood
conceptual terrain. They may however derail research if it turns out that they
are unrelated to the intrinsic meaning of what they representand people are
not aware or have forgotten that they are metaphors. Most physicists of the
present generation may not even understand the problem on hand since they
attribute concepts used in the construction of of a quantum theory erronously
with an intrinsic property of the construct; in the present concrete context
they are convinced that the classical string aspect in the classical N-G
action (and hence in the functional integral representation) must be somewhere
in the associated field theory. Since the commutator of the resulting string
field theory turns out to be pointlike localized they invent a string and
declare the point to be its center of mass. This mataphoric way of arguing is
supported by the historical fact that the quantization of pointlike classical
fields leads to pointlike quantum field. This coalescence of metaphoric and
intrinsic aspects was an extremely important event because it allowed QFT to
be dicovered before Wigner's intrinsic representation theoretical approach. It
carries the danger that people generalize it beyond its region of validity.

Precisely in order to avoid such mataphorical projections of classical aspects
into QT Heisenberg introduced the notion of "observable". The observable i.e.
intrinsic content of the N-G quantum string and its basic difference to
standard QFT is the abundance of the pointlike degrees of
freedom\footnote{According to my best knowledge the N-G string field is the
only generalized free field which permits a Lagrangian description. Whether
this makes this class of models physically more palatable (hitherto such
models were mainly used to restrict the postulates of QFT which do not allow
them).}. String theory has been around for more than four decades and it has
not only prepared the ground for the return of metaphors to quantum physics,
but it also has led to a loss of fundamental knowledge in particle physics.

In order to avoid any misunderstanding on this point and also not to appear as
a schoolmaster n the purity of of terminology in particle physics let me
emphasize that I am not criticizing the metaphoric and often very imprecise
terminology of particle physicists as compared to mathematicians per se. Most
physicists get perfectly over their surprise if they discover that Born's
famous article does not contain any x-space probability density nvolving
Schroedinger wave functions but is rather a paper on the notion of the cross
section in the Born approaximation (the $\left\vert \psi(x)\right\vert ^{2}$
density was introduced later by Pauli). Neither is he leaving his rocking
chair when he looks at Virasoro's original paper and finds neither the central
term nor an algebra with all frquency components. If string theory would be
just an inappropriate terminology for something well known or a cautious
terminology to avoid premature naive identification (as e.g. M-theory) one
could pass to business as usual..

Most of the great conceptual conquests of the post renormalization period in
QFT as the derivation of scattering theory and the related very subtle
connections between particles and fields have been reduced to computational
recipes. As a result these profound conceptual conquests are not passed on to
the younger generation. It is simply not true that history from the beginnings
of particle physics to string theory is a history of continuous progress. It
is not accidental that the rise of metaphoric thinking combine with a
"calculate and shut up" attitude coincides with the rise of string theory and
other physically unmotivated ideas. But this is not the case, the stringy
stuff wiggling in space is not just the stuff on which the Brian Green's Nova
Nova film project is based.but they rather enter the opening mantras of my
highly ....string-theory colleagues. In fact every string theorist I met and
conversed with firmely believes that his strings are one-dimensionally
extended objects in the same spacetime which serves as the living space of
pointlike fields.

The unusual and highly suspicious aspect of string theoretical matter as
compared to the Wigner classification of matter was already visible in the
dual resonance model \cite{Vec} in terms of a chiral conformal field theory.
Here the arena of the action of the Poincar\'{e} group is the \textit{target
space}, a very unusual situation indeed because the target space (a
metaphorical quantum analog to the field space of classical fields) is arena
of action for internal symmetries and these are usually given in terms of
compact groups. However for chiral theories the internal symmetry concepts are
less restrictive and it turns out that the kind of conformal theory behind the
dual resonance model can have a noncompact (Poincar\'{e}) symmetry but only if
the target space has 26/10 spacetime dimensions. But no matter whether the
Poincar\'{e} group acts on source or target space, modular localization, which
as emphasized before is always intrinsically related to the representation of
the Poincar\'{e} group, is the \textit{sovereign over quantum localization}
and not some classical string aspect of a N-G Lagrangian. This means in
particular that the classical string localization is irrelevant for the
quantum localization but certainly plays a role in the arrangement of the
irreducible components within a generalized free field.. The difference
between classical and quantum string localization can be sharpened and put
into the form of the following dictum: \textit{indecomposable} \textit{quantum
strings cannot be obtained from quantization and quantized classical strings
do not lead to quantum strings. }

The true physical content of the canonical quantization of the Nambu-Goto
Lagrangian is that of a generalized free field with infinitely many mass and
spin components\footnote{Usually the name generalized free field is reserved
for a fields whose c-number (anti)commutator has contribution from many
(possibly continuous) masses and spin.}. Such pointlike fields have different
properties from standard Lagrangian fields. Their phase space degrees of
freedom have a larger cardinality, and as a consequence they violate the
standard properties of thermal behavior (they have a Hagedorn temperature or
no temperature states at all) and also those of causal propagation. But apart
from the lack of causal propagation (time slice property, causal shadow
property) they satisfy all Wightman axioms, including spacelike commutativity.
Their existence has been (and still is) taken as an indication that the
Wightman axioms are too general and need further restriction\footnote{This was
noticed quite early in the history of QFT and let to the "time-slice
"requirement \cite{Ha-Sc}..}. Another case in which such an overpopulation of
degrees of freedom appears is the AdS-CFT correspondence in which e.g. a free
field on the AdS side is converted into a generalized free field \cite{Re} on
the conformal side.

Not all attempts to make physical sense of the quantum Nabu-Goto model with
the help of canonical quantization have ended in metaphors. There is however
an approach which avoids a canonical quantization by utilizing the fact that
the N-G model is classically integrable. In that case it seems to be
reasonable to find the classically conserved charges and their Poi$_{{}}$sson
bracket relations and (after verifying that there are no anomalies) to
quantize this algebraic structure \cite{Po}. In this case there is no problem
with re-parametrization invariance nor with locality since the conserved
charges are global reparametrization-invariant quantities and the construction
of the positive energy representations of their Poisson bracket structure
reinterpreted as an operator commutation structure is an eminent reasonable
procedure. However it is known that the content of such a theory is
inequivalent to the string theoretic quantization of the Nambu-Goto Lagrangian
\cite{Bahns}. The latter is driven by the aim to obtain a Lagrangian canonical
setting for the dual model and not by understanding the intrinsic quantum
content of the classical N-G equation viewed as a integrable model.

Related to the wrong metaphor of the N-G strings being little (or in more
recent times also large) stringy objects in the sense of localization in
spacetime\footnote{Unfortunately after having used erronous terminology for
more than 4 decades there is no way avoiding its continued use.} is the
enormous regression of string theory in all matters which are related to
spacetime localization concepts. One of the most impressive achievements after
the discovery of renormalized QED was the derivation of time-dependent
scattering theory and the closely connected improvement in the understanding
of the rather subtle connection between particles and fields. Problems which
appear insoluble in a pure S-matrix approach, as unitarity and
macro-causality, became linearized and hence manageable in the setting of
correlation functions of fields. The corresponding nonlinear properties for
the S-matrix which resisted attempts to implement them by "hand" are then
delegated to the (successful) proof of asymptotic convergence for large times.
Assuming that crossing holds for formfactors (a reasonable extrapolation of
what has been proven in QFT) though one cannot show the existence of an
associated QFT, at least its uniqueness can be established \cite{unique}.

String theory regresses on all these points, it remains a cooking recipe
without any concepts which could explain the validity of those properties
which are indispensable for any kind of particle physics as unitarity
macro-causality and crossing in the sense of section 4. If this regress would
be of a transitory nature on the way to something deeper which contains the
previous concepts on which the great physical successes of QFT is founded, one
could live with it for a limited amount of time; but are more than 4 decades
still a reasonable amount? The most devastating effect of string theory is
that it has led to a kind of (often arrogant) new type of particle physicists
who thinks what he learned under the label QFT as a preparation for string
theory suffices for doing particle physics. The fact that he cannot understand
structural nonperturbative results of QFT is of no concern because he accepted
the (blatantly wrong) message that QFT is, apart from some computational
details, a closed subject.

The real damage caused by the 4 decade lasting reign of string theory is not
that it leads to conceptual confusions and has not produced any physically
tangible result, but rather that is wiped out fundamental knowledge which will
slow down future progress in a post string era..

\section{TOE time, or particle theory in times of crisis}

Although among experts there is general agreement that particle physics is in
the midst of a crisis, not all share the optimism of some about a new
orientation coming out of the new generation of LHC experiments. Indeed it is
difficult to imagine that experiments can give new conceptuals directions in a
situation in which the experimental planning and the interpretation of
measured results depends metaphors as placeholders of principles. In addition
there ideas, as supersymmetry which are veterans in holding out against all
absence of any evidence. In its long history of more than 40 years and
thousands of publication there has been no no meaningful theoretically
consistent idea of a controllable breaking. Why such such a social construc
disappear with a whimper in the tunnels of the LHC? Certainly not because of
any new incompatibilities with observations; if against all odds it happens it
will be the result of exhaustion in finding excuses. Feynman saw the danger in
this observation-resistent and revolution-hardened return of metaphoric
reasoning when he commented on one occasion that string theory uses excuses
instead of arguments.

The diagnosis about the underlying causes varies widely according to age and
background. An often heard opinion is that the ascend of metaphoric ideas and
the increasing popularity of theories of everything\footnote{As far as I know
the first TOE came with the German name, it was the Heisenberg Weltformal (a
nonlinear spinor theory). Pauli supported it at the beginning but later (after
Feynman's criticism) turned against it. My later Brazilean collaborator
visited Munich at the end of the 50s and got so depressed about the circus
around this Weltformel that he had doubts about his decision to go into
particle physics. Fortunately that TOE remained a local event.} is the result
of stagnation in the mainstream of particle physics, i.e. that of QFT in
general and of the standard model in particular. The difficult and
time-consuming way to make genuine progress is not on par with people who are
after rapid success; they look wistfully at the beginning of the standard
model when it was possible to get progress with a relative modest amount of
knowledge\footnote{The discovery of QED two decades before was hard work
because all the renormalization technology and its conceptual basis had to be
developed.}, hardly any new concepts and no new formalism. It is much more
groovy to contemplate about a theory of everything than to labor with an
unfinished theory with more than 20 parameters, especially if the Zeitgeist
permits to make a good living by expanding the metaphors of a TOE.

It is true that many ideas which looked promising at the beginning (as e.g.
the unification through a confluence of the three couplings at a suitably high
energy) became frozen in time. But perhaps the stagnation of the main stream
of particle physics is itself already the result of deeming it unprofitable to
make any long time conceptual investment if a sufficiently large community
thinks that pursuing a TOE can bring rapid progress.

A theory whose intrinsic properties are unknown, apart from unresolved
metaphors, is a breeding ground of further-going metaphoric thinking. This is
particularly evident in the application of string theory to what its admirers
consider its central achievement: the famous AdS-CFT holography. This is an
exemplary case were one metaphor begs the next one, and as such it is
extremely informative for the points raised in this essay; so let us look at
this issue with some care.

Already in the 60s the observation that the 15-parametric conformal symmetry
which is shared between the conformal of 3+1-dimensional compactified
Minkowski spacetime and the 5-dim. Anti-de-Sitter (the negative constant
curvature relative of the cosmologically important de Sitter spacetime)
brought a possible field theoretic relation between these theories into the
foreground; in fact Fronsdal \cite{Fron} suspected that QFTs on both
spacetimes share more than the spacetime symmetry groups. But the modular
localization theory which could convert the shared group symmetry into a
relation between two \textit{different spacetime ordering devices} for the
\textit{same abstract quantum matter substrate} was not yet in place at that
time. Over several decades the main use of the AdS solution (without its
covering manifold), similar to Goedel's cosmological model with self-closing
timelike worldlines, has been to show that Einstein-Hilbert field equations
besides the many desired solution (as the Robertson-Walker cosmological models
and the closely related de Sitter spacetime) also admit unphysical solutions
which lead to time machines, wormholes etc., and therefore should be further restricted.

The AdS spacetime lost this purpose of only providing counterexamples and
began to play an apparently more constructive role in particle physics when
the string theorist placed it into the center of a conjecture about a
correspondence between a particular maximally supersymmetric massless
conformally covariant Yang-Mills model in d=1+3 and a supersymmetric
gravitational model. The first paper was by J. Maldacena \cite{Ma} who started
from a particular compactification of 10-dim. superstring theory, with 5
uncompactified coordinates forming the AdS spacetime. Since the mathematics as
well as the conceptual structure of string theory is poorly understood, the
string side was identified with one of the supersymmetric gravity models
which, in spite of its being non-renormalizable, admitted a more manageable
Lagrangian formulation and was expected to have a similar particle content as
the less understood superstring theory from which it originated. On the side
of CFT Maldacena placed a maximally supersymmetric gauge theory of which
calculations, which verify the vanishing of the low order beta function,
already existed. The vanishing of the beta-function is certainly a
\textit{necessary} prerequisite for conformal invariance.\ The arguments
involved perturbation theory and additional less controllable approximations.

The more than 5.000 follow up papers on this subject did essentially not
change the status of the conjecture. But it was the kind of sociological
backup which elevated the Maldacena conjecture the most important result of
string theory and its putative connection with the still elusive quantum gravity.

The conceptual situation became somewhat more palatable after Witten
\cite{Witten} and Polyakov et al \cite{Polya} exemplified the ideas in the
field theoretic context of a $\Phi^{4}$ coupling on AdS using a D-dimensional
Euclidean functional integral setting, thus placing it into a form which is
closer to the old Frondsdal setting. Of course unlike the supersymmetric
Yang-Mills theory this self-coupled model is not expected to lead to a
conformal theory as the result of a trace anomaly for the energy-stress
tensor. The correspondence argument for the self-coupled scalar field ignores
this anomaly and consists in subjecting the conformally invariant building
blocks namely the propagator and the pointlike vertices to the requirements of
the AdS-CFT holography.

The model-independent \textit{structural properties of the AdS-CFT
correspondence} came out very clearly in Rehren's \cite{Rehren}
\textit{algebraic holography}. $\ $The setting of local quantum physics (LQP)
is particularly suited for questions in which one theory is assumed as given
and one wants to construct its corresponding model on another spacetime. Using
methods of local quantum physics one can solve such problems of isomorphisms
between models in a purely structural way i.e. without being forced to
actually construct a model on either side of the correspondence.

Since generating pointlike fields are coordinatizations of spacetime-indexed
operator algebras and, as with numerical valued coordinates in geometry, such
coordinatizations are highly nonunique and certainly not intrinsic an
algebraic formulation of a correspondence is more appropriate. For certain
structural questions i.e. whether the inverse scattering problem for an
S-matrix with the crossing property has a unique QFT to which it is the
S-matrix the algebraic formulation is the only meaningful one. \ It is not
surprising that holographic changes of spacetime encodings are more tricky if
expressed in terms of relations between pointlike fields. For example a
standard pointlike quantum field on AdS\footnote{Here "standard" means
originating from a Lagrangian or, in more intrinsic terms, fulfilling the
time-slice property of causal propagation. A free field is standard in this
sense, a generalized free field with an increasing Kallen-Lehmann spectral
function fails to have this property.} has very non-standard behavior on the
conformal side; there is an overabundance of degrees of freedom which is of
course what one expects of a correspondence in which a collection of degrees
of freedom which was natural in 5 spacetime dimension is squeezed into 4
dimensions. This can be nicely illustrated in case of a free AdS field which
under the correspondence becomes a generalized free field with a continuous
distribution of masses which \ carries the anomalous dimension of the
conformal generalized free field \cite{Du-Re}.

Here some informative historical remarks about generalized free fields are in
order. They were introduced in the late 50s by W. Greenberg and their useful
purpose was (similar to AdS in classical gravity) to \textit{test the physical
soundness of axioms of QFT} in the sense that if a system of axioms allowed
such solutions which appeared unphysical, it needed to be further restricted
\cite{Ha-Sc} (in that case the so-called causal completion or time-slice
property excluded generalized free fields). The unphysical aspect of the
generalized free field consisted in the breakdown of the causal shadow
property i.e. the operator algebra in a spacetime region generated by certain
generalized free fields are much smaller than the operator algebras of their
causal completions. Another related failure is the existence of a limiting
temperature (the Hagedorn temperature) or even worse, the nonexistence of
temperature states altogether. The fact that there have been many papers in
string theory about systems with the Hagedorn temperature does not mean that
nature has become more lenient with respect to older physical principles. Not
anything which originates these days from a physicist is "physical"; we are
not yet living in a Tegmark world \cite{Teg} where every mathematical belch
finds its physical realization in some universe of his conceived multiverse.

In the opposite direction the degrees of freedom of a "normal" CFT become
"diluted" on AdS. There are not sufficient degrees of freedom for arriving at
nontrivial compactly localized operators, the cardinality of degrees of
freedom is only sufficient to furnish noncompact regions as AdS wedges with
nontrivial operators, whereas the compactly localized double cone algebras
remain trivial (multiples of the identity). In the setting based on fields
this means that the restriction on testfunction spaces is so severe that
pointlike field $A_{AdS}(x)$ at interior points $x\in intAdS$ do not exist in
the standard sense as operator-valued distributions on Schwarz spaces. They
exist on much smaller test function spaces which contain no functions with
compact localizations.

Rehren's structural analysis adapted to the functional setting in order to
allow a comparison was dismissed by string theorists. Comparing the formal
transcription of Rehren%
\'{}%
s approach to a functional integral setting it was indeed difficult to see any
relation with Witten%
\'{}%
s functional integral treatment. But thanks to a functional identity
(explained in the Duetsch-Rehren paper) which shows that fixing functional
sources on a boundary and forcing the field values to take on a boundary value
via delta function in the functional field space leads under certain
conditions to the same result. In this way the apparent disparity disappeared
\cite{Du-Re2} \cite{ReLec} and there is only one AdS-CFT correspondence within
QFT and not two (namely that coming from string theory and that established by
Rehren's) as claimed by string theorists.

Theorems derived with mathematical rigor in a conceptually clear setting mean
nothing to people who are convinced to hold a TOE in their hand. Predictably
string theorists ignored Rehren%
\'{}%
s work even though it was now clear that there was only one correspondence; in
those few cases when they mentioned it after being questioned, they expressed
their disdain by labelling it as the "German AdS-CFT correspondence". Whereas
at the time of Feynman they at least had to think about excuses, the increase
of power and glory in the meantime has changed that behavior in the sense that
has come down to labelling.

As a veteran of QFT one finds oneself asking the question how did one get into
this strange movie? Her one depends on guesses. Excluding personal
animosities, the only question which comes to ones mind is whether there could
be anything in Rehren%
\'{}%
s rigorous results which comes into conflict with string ideology. Is it
perhaps the before mentioned mismatch between degrees of freedom as compared
to the string theorists conjecture that the correspondence should relate two
Lagrangian QFT? Maybe this is viewed as a stumbling block on the way to a TOE.
The reason why string theorist are unable to see the agglomeration of too many
degrees of freedom on the conformal side may be that the uncontrollable
approximations they use inevitably also thin out degrees of freedom; perhaps
there exists even a meaningful procedure to prepare a thinning out on the AdS
side so that the conformal theory will be of the normal kind as we know it
from conformal limits of Lagrangian theories (critical limits in universality
classes). Unfortunately the run after a TOE leaves such interesting questions
on the wayside.

There is however one deeply worrisome aspect of this whole development. Never
before in the history of particle physics have there been around 5.000
publication with inconclusive results on such a rather narrow subject. In fact
even nowadays, one decade after this gold-digger's rush about the Maldacena
AdS-CFT correspondence started, there is still a sizable number of papers
every month by people looking for nuggets at the same place but without
bringing Maldacena's gravity-gauge conjecture any closer to a resolution.

Even with making all the allowances in comparison with earlier fashionable
ideas in particle theory, this phenomenon is too overwhelming order to be
overlooked. Independent of its significance for particle physics and the way
it will end, the understanding of what really went on and it was presented by
the media will be challenging problem to historians and philosophers of
science in years to come. The main stimulus for this work is the hope that an
article like this could facilitate their extremely difficult work. Experience
with conjectures in particle physics suggests that a claim which remains
unproven for 10 years will never be proven. although after the passing of some
time it will be , depending on expediency, presented as an established fact.
In physics there are no Fermat-like conjectures which, after lying dormant for
a long time, are proved or disproved.

Since commentaries like this run the risk of being misunderstood, let me make
perfectly clear that particle physics was a speculative subject and it is
important that it remains this way. Therefore I have no problem whatsoever
with Maldacena's paper; it is in the best tradition of particle physics which
was always a delicate blend of a highly imaginative and innovative
contribution from one author with profoundly critical analysis of others. I am
worried about the loss of this balance.

My criticism is also not directed against the thousands of authors who have
entered this area in good faith believing that they are working at an
epoch-forming paradigmatic problem because their peers gave them this
impression. A phenomenon which represents the Zeitgeist cannot be pinned to
particular persons; this is a theme to which we will return in the last section.

The field theoretic AdS-CFT correspondence is a special case of a holographic
projection which maps a QFT onto a lower dimensional QFT on its boundary. The
most useful kind of holography is that of a causally closed theory in a bulk
to that of its \textit{causal boundary}. The best studied case is that of a
wedge localized algebra and the algebra on its (upper) causal boundary which
constitutes half of the lightfront. This algebraic holography uses modular
theory which requires to work with localized algebras. Of course once one
knows the holographic projection on half the lightfront it is easy to
reconstruct the full lightfront algebra. The algebraic lightfront holography
can be used to derive a universal area law for the localization entropy
associated with a horizon \cite{add}.

Another implementation called projective holography consists in applying the
holographic projection directly to pointlike fields \cite{ReLec}. This
holography on null surfaces is extremely useful because, different from the
AdS-CFT holography, it is not a correspondence but rather a projection, in
fact the image on the null surface is much simpler than the bulk. This makes
it possible to study certain properties which in the bulk would be
inaccessible with the present technology.

Even though the bulk algebra is not uniquely determined by its holographic
projection, one could hope that with additional assumptions the reconstruction
of the bulk could be unique. This is certainly the case if the holography
permits a description in terms of generators whose ambient Poincar\'{e}
transformation properties are known. Interesting nontrivial examples are
provided by factorizing models in d=1+1 \cite{contro}\cite{Hol}. Different
from the AdS-CFT correspondence\footnote{In geometric terms this is a
holography between the full AdS bulk and its timelike brane at infinity. All
other studied holographies are onto lightlike boundaries.} the lightfront
holography has no mismatch between degrees of freedom since as a result of the
holographic projection there is a thinning out of degrees of freedom, i.e.
their density on the lightfront is natural in terms of the lower dimensional
lightfront submanifold.

Whereas the null surface holography admits rich applications, is less clear
what the AdS-CFT could offer to a theorist who does not believe in the holy
Grail of string theory. For this it is helpful to remind oneself of the
intrinsic conceptual meaning of holography. One starts from a theory which
consists of an abstract algebraic substrate (examples: the CCR or CAR
algebras) which is structured with spacetime as an ordering device (in the
sense of Leibniz) so that for each (without loss of generality causally
closed) spacetime region one has a subalgebra of the global algebra. There is
nothing more to a QFT than this spacetime-indexed "net of algebras"; among
particle physicists with a profound knowledge of QFT it is well known that the
spatial ordering of the algebraic substrate is all there is i.e. every
property of quantum matter can be derived from this picture\footnote{In fact
there is even a more general (and more abstract) characterization of QFTs in
that a net of algebras including its spacetime ordering (and the action of the
Poincare group) is uniquely determined in terms of the relative positioning of
a finite number of "monads" (copies of the unique hyperfinite type III$_{1}%
~$factor algebras) in a common Hilbert space.}. In this setting holography is
a rather radical change of the spacetime ordering device keeping the same
material substrate.

From this point of view one potentially interesting application comes to ones
mind which is not so different from the lightfront holography: simplification
of certain aspects of a conformal QFT (e.g. a supersymmetric Yang-Mills model)
in the AdS perspective; despite many interesting analogies between chiral
theories and higher dimensional QFT \cite{To} little is known about
higher-dimensional conformal QFTs. For this purpose the interpretation on the
AdS side is important, what matters is that there are some simplifications
i.e. one could be at a lookout for integrable substructures which were too
much hidden in the original spacetime ordering. Since such investigations
would also be independent on the degree's of freedom issue this would be a
completely incontrovertible meeting ground with string theorists who already
started such investigations. In principle such changes of spacetime encoding
in order to improve computational accessibility make sense without conformal
covariance\footnote{Most properties of particle physics are lost in the
conformal critical limit, the only candidates which are expected to survive
such limits are highly inclusive cross sections.}.

Let me emphasize again that I believe that holography is a technical tool and
not a physical principle. It simplifies certain aspects of a QFT at the
expense of others (i.e. it cannot achieve miracles). The use of such ideas in
intermediate steps may have some technical merits, which is quite a lot in an
area where credible nonperturbative statements are hard to come by.

String theory is the first theory (perhaps only the first after the phlogiston
theory of burning) which despite missing observational as well as
mathematical/conceptual credentials got firmly entrenched in the mainstream of
particle physics. The question is how was this possible in a science which is
considered to be the home of rationality and observability is a difficult one.
Human activities even in the exact sciences are not completely independent of
the Zeitgeist and the glory and power of the invigorated post USSR capitalism
with its "end of history" frame of mind at the begin of a new millennium was
looking for some shiny scientific counterpart whose glory was not clouded by
having 20 unexplained parameters around. It was nevertheless the easy and even
somewhat (theoretically) unmerited of the standard model which contributed to
that somewhat arrogant frame of mind to go all out for a TOE.

String theory was the appropriate vehicle since its mathematics is relatively
rich so there was no danger to fall back behind the mathematical
sophistication which since the time of Atiyah and Witten dominated parts of
QFT. Although there were (and still are as we have seen) many metaphoric
aspects concerning its physical-conceptual content, it is certainly less
vulnerable to criticism that the previous failed two attempts of a
"Weltformel" and of a unique S-matrix bootstrap.

The lack of challenging pronouncements of string theory is rarely the result
of fear to damage one's reputation and career although with leading and
influential particle physicists enthusiastically supporting this certainly
plays a role. But what weighs more heavy here is the fact that the profound
knowledge which would be necessary to start a different direction is not
passed on; the limited amount of time one has for carving out a career in
particle physics excludes the acquisition of potentially important knowledge
which may be important for innovative work but which happens not to be in the
vicinity of one's line of research. This is of course part of a much wider
sociological problem but it is aggravated by string theory since by covering
up this problematic it makes it even virtually impossible that anybody
acquires that knowledge which could lead to its own demise. Experimental
results as those expected from the LHC are hardly capable to bring down string
theory. There are simply too many excuses in case of e.g. of a continued trend
of no signal for supersymmetry. String theory may be poor in predictions, but
it is rich in post-dictions.

\section{Responsibilities? No scapegoats}

In almost all areas of human activities times of crisis are not only times of
trying out new directions but also times of lookouts for culprits and
scapegoats. Given the necessarily highly speculative nature of particle
physics research on fundamental theoretical problems it is self-defeating to
give space to the latter activity. It would be a very bad idea to curb the
speculative side and restrict particle theory to mathematically controllable publications

In normal times proposals which do not comply with existing theories often led
to new physics; and in order to find out whether a proposal has such a
potential one needs some time. To be accepted by the community and be added to
the pantheon of physics, a theory must be subject to observational tests and
undergo a critical review of its physical conceptual content which certainly
includes a clear positioning with respect to the established theory.

String theory was the first such proposal which, as the result of its Planck
scale interpretation, was exempt from the observational requirement. In this
situation of lack of direct observability, a fundamental theoretical
discussion about its conceptual basis would have been of the highest
importance and priority. At the time of its existence as phenomenological dual
model proposal for strong interaction, there was yet no compelling reason to
do this, but when it laid claims to be the first TOE which incorporates
gravity, there should have been an extensive discussion. This chance was
missed; unlike the similar situation around the S-matrix bootstrap during the
60s, there was no such discussion when the dual resonance model changed to
string theory. String theory moved directly from strong interaction
phenomenology to a TOE by only adjusting the numerical string tension from its
value in the Regge setting to the Planck setting which amounts to a jump of 15
orders of magnitude without changing one jota in the formalism..

Our main criticism of string theory in section 6 was a refutation of its
metaphoric trespassing of Heisenberg's dictum on quantum observables i.e. the
insistence in attributing to quantum objects an intrinsic interpretation which
is not only independent of classical analogies but also the quantization
method by which the object has been constructed. Contrary to general opinion
and to terminology, the objects associated with the canonically quantized
Nambu-Goto Lagrangian are point-localized generalized free fields and there is
no reason whatsoever to expect that interactions will magically convert them
into string-localized objects\footnote{Such unfounded expectations of string
theorists irritated Feynman and led to his statement that string theorists
offer excuses instead of explanations.}. Unfortunately the physical content of
string theory hinges on the string interpretation.

In this context I cannot resist to cite an aphorism (of unknown source), which
perfectly characterizes the problem with metaphors in QFT : \textbf{something
which looks, moves, smells and sounds like an elephant is really an elephant;
in QFT there is no recourse to metaphors which permit different conclusions.
}For the case at hand: the N-G quantum object is really a generalized free
field despite the fact that its classical action in a functional integral
representation suggests a string-like localization. Pointlike fields with an
abundance of particles (infinite mass- and spin-towers) do not arise from
Lagrangians in QFT, but the N-G Lagrangian does not arise in QFT. Nevertheless
one has all reasons to be surprised about obtaining a generalized free
field\footnote{Generalized free fields were hitherto only used in order to
argue that settings which admit them (as e.g. the Wightman setting) need to be
further restricted. Phases space restrictions on the cardinality of degrees of
freedom certainly eliminate such fields which lead to abnormal thermal
behavior (appearance of a Hagedorn temperature). They have never been observed
in nature but do occur in theoretical constructs as e.g. the correspondence of
a free AdS field to its conformal lower dimensional image.}

In the concrete application: a relativistic quantum theory whose commutators
signal pointlike localization is really a pointlike localized QFT even if it
carries many more degrees of freedom than a standard (Lagrangian, Wightman)
QFT model.

In order to avoid any misunderstanding on this point let me emphasize that I
am not criticizing the somewhat metaphoric style of research of theoretical
physicists as opposed to mathematicians. Most physicists can perfectly live
with the realization that the so-called Born probability density
interpretation of the square of $\left\vert \psi(x)\right\vert ^{2}$ cannot be
found in Born's paper which rather deals with the notion of the scattering
cross section in the Born approximation. He also will not leave his rocking
chair if he notices that the famous paper by Virasoro does neither contain the
central term nor all the frequencies (he only wrote the algebra for the
parabolic subgroup). If the string in string theory would be just a misleading
name and everybody would know that the N-G Lagrangian leads to a very special
generalized free field and hence the expectation for the interacting case
would be a pointlike field with an overpopulated quantum field theoretical
phase space the only raised eyebrows would be those of schoolmasters. In that
case the way for an intrinsic understanding of the position of the
string-induced generalized fields within the enormously large class of
pointlike APD fields (abundant phase space degrees) would have been free. By
following the autonomous logic of this problem one had (and still has) a
chance of a deeper understanding of the the metaphoric idea of \textit{a}
\textit{unitary} \textit{Poincar\'{e} group representation on a fluctuating
target space of a chiral QFT} and its hidden power to tell string theory on
what spacetime dimension it has to live.

But this is unfortunately not how things developed; the stringy objects
wiggling in spacetime have not been limited to the stuff from which Brian
Green composes his Nova videos, but they are rather the core of opening mantra
of talkers about string theory. And let me add, every string physicist I
talked to believed that string theory describes precisely such objects.

The return to a past which was less strict on observability (at least at the
level of a thought experiment) has made a long lasting imprint on particle
physics because a misleading metaphor on such a central place once accepted by
leading researchers is bound to spread and contaminate the whole setting.
Indeed, accepting the first metaphor leads to the second namely defining
interaction through Euclidean "tubism" and so on, up to the point where the
whole setting becomes incomprehensible in any autonomous sense. It seems that
in this way one becomes well-prepared for the ultimate metaphoric step; the
fundamentalist idea of a TOE.

A criticism of its central claim of replacing point- by string- localization
as in this essay (but formulated at the right time in the 60s), may not by
itself have changed much, but it could at have least prevented the metaphoric
excess concerning terminology.

Now, almost 4 decades later, and in spite of worldwide attention and an
enormous number of publications, the conceptual status of string theory has
remained as obscure as at the beginning. This conceptual uncertainty and in
particular the above central metaphor about its localization is the ideal
breeding ground for further metaphorical thoughts (even in cases in which
completely rational results would be available. One such idea, namely the
string theoretical view of the AdS-CFT correspondence and the more than 5.000
publications without credible results was already subjected to a critical
review in the previous section.

The present critical situation in particle physics is regrettable, since many
of the string theorists are very competent and intelligent people, well
prepared to make important observations. Without Maldacena%
\'{}%
s contribution the issue of a field theoretic AdS-CFT correspondence would not
have arisen and we would still be at the level Fronsdal's group theoretic
observation. Even to people (as myself) who are not able to attribute to it
the status of a holy Grail in the string theoretical setting of a TOE
containing quantum gravitation, the AdS-CFT correspondence is an interesting
observation because it makes one aware that a radical change of the spacetime
ordering device brings about a drastic change of the physics, even if the
shared abstract matter substrate is "only" being re-organized in spacetime.

It is regrettable that following the fundamentalism of a TOE seems to make
people intellectually arrogant and blind or immune against criticism. An essay
like this has hardly any effect on the string community; it would be utterly
naive to have illusions on this point. Its only value may be that it
facilitates the task of future historians and philosophers of science to
understand what happened to particle physics at the end of the millennium. One
can be sure that there will be a lot of public interest in work dedicated to
obtain some explanations about this strange episode in the midst of the exact sciences.

In the case of the S-matrix bootstrap which preceded string theory (and from
which it inherited the idea that the central object should be the S-matrix
with crossing property which it then changed to duality), there was
substantial criticism\cite{Jost}. It was not directed against the S-matrix
properties as such after all had been abstracted from QFT. Rather the main
cause of irritation came from the anti-QFT ideology behind the S-matrix
supremacy and the claim of the uniqueness of the solution to those principles
(a TOE without gravity). If one removes these excessive claims, the bootstrap
framework consists of a list of correct S-requirements, but unfortunately
(apart from 2-dim. factorizing models) provides no means for their implementation.

The S-matrix community actually enriched particle physics by making quantum
field theorists more aware of the principle of \textit{nuclear democracy}
which is the on-shell relic of \textit{Murphy's principle of interacting QFT}
namely that \textit{all states which can be coupled according to their
superselected quantum number are actually coupled}\footnote{I apologize to the
reader for using metaphoric terminology, but in this case I do not know any
better formulation which would be suitable in an essay like this.}. The
on-shell version of this principle is called the principle of \textit{nuclear
democracy}. It is beautifully realized in the explicitely solved factorizing
models. Contrary to a widespread opinion confinement does not contradict
nuclear democracy since invisible/confined/dark objects may be associated with
semiinfinite string localized fields which cannot be produced by ordinary
matter but live nevertheless in a common physical Hilbert space non all
fundamental may appear on-shell and lead to compactly localized composites
which generate ordinary matter.

To say it again, it would be wrong to attribute the demise of S-matrix
bootstrap at the time of the strong return of QFT in form of QCD and the
discovery of the standard model to errors or shortcomings in the S-matrix
postulates. The fierce backlash was rather caused by the loud-mouth attempts
of some of its leading defenders to cleanse particle physics from QFT in
conjunction with the poverty of credible results.

A convincing argument for the soundness of the underlying S-matrix principles
came from the success of the bootstrap program and its extension to the
bootstrap-formfactor setting in the context of factorizing models which
started in the late 70s. Of course the existence of infinitely many such
models for which the classification of bootstrap S-matrix solutions
contradicted totally the TOE expectations; perhaps for this reason did not
interest the bootstrap protagonists as Chew, Mandelstam, Stapp. They were
after a TOE in the setting of the S-matrix bootstrap and the least they would
be interested in (except Mandelstam) was a method for a new nonperturbative
constructive setting of factorizing QFT using formfactors of local observables.

The sociological situation at the crucial time of the transmutation of the
dual model to the string-theory TOE was quite different. Even though QFT had a
strong come back and the standard model was universally hailed as impressive
(and from a modern perspective somewhat unmerited and
untimely\footnote{Whereas one may have serious doubts about the validity of
the well known aphorism about string theory being a theory of the 21st century
which fell by look onto the 20. century, its adaptation to the standard model
appears more credible.}) achievement, for many it did not present a large
enough projection screen of their ambitious imagination, in particular because
it said nothing about gravity which moved increasingly into the center of
interest. There were two ways to pursue such an aim, either follow the
reasonably safe but somewhat lengthy and winding path of QFT in CST, hoping
that eventually one will get to a point which requires the natural
intervention of quantum gravity. The other one followed by the majority was to
play a riskier game by trying out more speculative attempts which, if
successful, could lead to quantum gravity and possibly the holy Grail of a TOE
in what alpinists call a diretissima.

The first path had led to unexpected fundamental insights. Particle physicists
became aware of thermal aspects of causal localization which were first noted
in a rather abstract mathematical context of QFT\footnote{The thermal KMS
property is a byproduct of the application of modular operator theory to QFT
.\cite{Bi-Wi}.} and independently in the more concrete physical context of
black holes localization on bot sides of an event horizon. Extensive
investigations, with started with the covariance properties of the
energy-stress tensor in curved spacetime, finally led to the discovery of the
\textit{local covariance principle} which states that isometric spacetime
submanifolds have unitarily equivalent quantum physics, thus this principle
goes a long way towards establishing properties in QFT in CST which strictly
speaking were hitherto ascribed to quantum gravity.

There are also recent results about cosmology. In particular there is
presently no reason to speak about a cosmological constant trouble since there
exists at present there exists no calculation which is consistent with this
new principle (the quantum mechanical estimate which led to the well-known
much too large value is not consistent).

On the other hand the string theory approach has led to many computations but
no physical insight apart from metaphoric scenarios. It is the living
demonstration of the "compute and shut up" maxim; whoever wants to see
particle theory as an unfolding of principles or at least as a conceptional
guide (say on the level of the particle physics discussions of the 60s and
70s) will be disappointed.

In view of the fact that a highly speculative task as research in particle
theory depends on a delicate calibration between the inventive and the
critical side of fundamental research, the present situation is worrisome.
Even having witnessed the beginnings of string theory one looks in vain for a
scientific explanation how a failed phenomenological model coming from the
setting of Regge poles could have been elevated by not much more than a
\ "semantic trick" (and a scale-sliding of 15 orders of magnitude) to a
trans-Planckian TOE, a truly unique phenomenon in physics which would dwarf
any achievement of Galileio, Newton, Einstein, Heisenberg and you name it.
There is really nothing which can match the grandeur of superstring theory.

Just imagine for a moment that a particle physicist of the 60s would be
time-shifted into the millennium string theory community. The only aspect
which would prevent him to think that he is in a madhouse for particle
physicists is the presence of some Nobel prize winners. He would listen to
praises of string theory as "the gift to the 21 century which fell by luck
already into the 20$^{th}$ century".and note with disbelieving amazement that
the depth of this theory requires still several decades before one knows what
is in it. He also would have to get used to a new deployment of big Latin
letters. To him the use of the letter M with the invitation to interpret it as
mystic, magic etc. a plea for the use of metaphors in particle theory
research. Our hypothetical time-shifted visitor will also still somewhat
disbelievingly listen to statements about attempts to make sense of new
developments as the following (taken from Wikipedia):

\textit{In the 1990s, Edward Witten and others found strong evidence that the
different superstring theories were different limits of an unknown
11-dimensional theory called M-theory. These discoveries sparked the second
superstring revolution. When Witten named M-theory, he didn't specify what the
\textquotedblright M\textquotedblright\ stood for, presumably because he
didn't feel he had the right to name a theory which he hadn't been able to
fully describe. Guessing what the \textquotedblright M\textquotedblright%
\ stands for has become a kind of game among theoretical physicists. The
\textquotedblright M\textquotedblright\ sometimes is said to stand for
Mystery, or Magic, or Mother. More serious suggestions include Matrix or
Membrane. Sheldon Glashow has noted that the \textquotedblright
M\textquotedblright\ might be an upside down \textquotedblright
W\textquotedblright, standing for Witten. Others have suggested that the
\textquotedblright M\textquotedblright\ in M-theory should stand for Missing,
Monstrous or even Murky. According to Witten himself, as quoted in the PBS
documentary based on Brian Greene's \textquotedblright The Elegant
Universe\textquotedblright, the \textquotedblright M\textquotedblright\ in
M-theory stands for \textquotedblright magic, mystery, or matrix according to
taste.\textquotedblright}

This playful name-coquetry permits the writer of these lines to avoid to take
a stand on the content. Needless to add that the opening mantra for all
introductory talks/articles on string theory usually starts with something
like this:

\textit{String theory is a model of fundamental physics whose building blocks
are one-dimensional extended objects (strings) rather than the
zero-dimensional points (particles)}

In recent times the emphasis on the stringlike localization of their objects
as opposed to pointlike has somewhat subsided, but it would probably be too
optimistic to think that this could have something to do with a better
appreciation of the intrinsic meaning of localization of quantum objects, one
of the most subtle in particle physics. To hope for a statement which corrects
4 decades of incorrect narration about string localization will be in vain; if
it happens at all, then only as part of a swan song of string theory.

Of course metaphors in quantum theory were used before, but they only appeared
in connection with intuitive arguments were it was clear that they were
transitory placeholder i.e. represented an temporary formulation which should
be replaced by a intrinsic permanent one.

There is a certain grain of (perverse) truth in string theorists self-defense
in hard pressed situations (e.g. panel discussions or interviews) when they
adopt the argument of David Gross that notwithstanding all criticism,
superstring theory is "the only game in town". Similar to the words of a
character in the late Kurt Vonnegut's short story (which Peter Woit \cite{Wo}
used in a similar context):

\textit{A guy with the gambling sickness loses his shirt every night in a}

\textit{poker game. Somebody tells him that the game is crooked, rigged}

\textit{to send him to the poorhouse. And he says, haggardly, I know, I}

\textit{know. But its the only game in town.}

Kurt Vonnegut, The Only Game in Town \textit{\cite{Vo}}

the situation in string theory is self-inflicted, although its defenders make
it appear as the result of an inevitable development in particle physics.
Self-fulfilling prophesies with disastrous consequences are better known in
politics, but it appears that with the widespread acceptance of string theory
this is the first time they also entered particle theory although the meaning
of "disastrous" in both cases has different dimensions

I do think that pied pipers with their TOE tune have a negative influence on
particle physics. They tend to direct the young ambitious minds away from the
main line of particle research in which an increasing unification was obtained
as a side result of a natural step for step natural development following the
intrinsic logic and avoiding jumps which create gaping conceptual holes as the
ones which led to string theory.

The new fundamentalist style can only be maintained within a very large
\ worldwide group as part of "big science". Just imagine that somebody would
use that metaphoric way of arguing without having a large community behind
him. The fear that people could think that he must be off his head would
probably prevent him. The metaphoric approach aiming at a TOE within a large
TOE receptive community and big science are the prerequisites for sustaining
superstring theory. In fact once a bright individual has entered that area he
already has accepted the use of metaphoric arguments and he probably will not
even become aware that he is moving outside Heisenberg's notion of quantum
observables which bans the use of classical metaphors. prohibits the use of
that deep contradiction with the fact that quantum theory started with
Heisenberg's banning metaphorical classical arguments in favor of strict
adherence to quantum observables.

For more than three decades considerable intellectual and material resources
in the form of funding research laboratories and university institutions have
been going into the advancement of string theory. and this has led to a
marginalization of other promising areas. So the statement that there is no
other game in town has becomes true in a very literal material sense. At no
time before has a proposal, which for more than four decades did not
contribute any conceptual enrichment or observable prediction to particle
physics, been propagandized even into the most remote corners.

In front of the scientifically interested public it is not easy these days to
raise doubts about the scientific soundness of string theory and extra
dimension, because as a traditional scientist one is not accustomed to
confront propaganda a la Green \cite{nova}\cite{Green} to which the science
interested public is exposed. But inasmuch as in a democratic society it is
difficult to blame individuals for deformations, the miscarriage of particle
physics can only take place if the educated public, meaning we, starts
tolerating or accepting arguments uncritically and allows scientific spin
doctors to use the media in order to praise their favorite product. A
successful seduction does not only require a skilled seducer, but also people
who despite all their knowledge and critical capacities permit themselves to
be seduced which is not a good situation for starting a blame game. As
political events show, these processes of mass seduction only happen under
special circumstances i.e. they are strongly coupled to the Zeitgeist whose
direction in particle physics was for a long time coming that of a TOE as a
contribution to the millenniums power and glory. This raises the hope of a
future change, but unfortunately such long term derailments come at the prize
of a lot of destruction which in particle physics amounts to a loss of knowledge.

The long term damage which string theory is causing is not just resulting from
a distorted picture of particle physics, as was described in this essay.
Highly speculative subjects as particle physics develop in a delicate balance
between innovation and critique and are therefore more prone to derailment.
The ultraviolet catastrophe and the S-matrix bootstrap were two of the bigger
pre-string transitory derailments which dominated the scene for not much more
than a decade and they did not led to changes in the way basic courses were
structured. So the future physicists having acquired unbiased well-balanced
knowledge about particle physics, unlike the above Vonnegut character, never
had to confront a "no other game in town" situation as a result of lack of
knowledge or misinformation. This situation has changed completely with string
theory being the hegemon of particle physics. Large parts of the central
concepts of QFT as scattering theory, the subtle connection of particles and
fields with localization, i.e. in particular those topics which were explained
in this essay are not taught any more. Instead topics as e.g. Calabi-Yao,
moduli,..which are related with functional representations and have less
direct relation\ to particle physics have replaced them. Since string theory
is around for more than four decades it will be extremely difficult to find
one's way out which in turn will lead to a prolongation of the crisis.

Combatting the propaganda coming from string theorists and presenting the
negative effects of the sociological supremacy of the string theory community
is the intention of two recent books \cite{Wo}\cite{Sm}. Although I would not
underwrite each of their arguments I do broadly with their conclusions. In
fact the avoidance of being repetitious was one argument for rewriting the
present essay in the direction of theoretical consistency problems.

Having stated my viewpoint, the reader will not be surprised that I have no
sympathies with TOEs (theories of everything) as a kind of final solution of
particle physics although natural gradual unifications as they happened in the
past are very desirable. The idea that Einstein's Dear Lord permits some
string theorist to find a closure to fundamental physics, so that for the rest
of all days the curiosity of humanity about its material world will end in
intellectual boredom (or at best can be re-directed into filling in some
technical details, finding new applications or preparing something for the
entertainment industry as Brian Green"s cinematic production \cite{nova})
appears to me outright quixotic. Ideas like this probably will be cited by
historians in a more distant future as representing the hubris and
intellectual (not necessarily personal) arrogance of a past millennium
Zeitgeist. I hope that if particle physics is not able to find its way back to
modesty by its own, that the development and astrophysics and cosmology will
help to prod it away from its present metaphoric path.

The perhaps most blunt attacks against the collateral damage of string theory
on the basis of the collateral damage it causes came from condensed matter
physics notably from the Nobel laureate Robert Laughlin \cite{Laugh}. If the
fundamentalist TOE setting needs an extreme counterpart, it is the radical
form of the emergence principle as defended by Laughlin. It is hard to agree
with an underlying philosophy which negates even those forms of natural forms
unification which were very successful in pre TOE times. Even in very
successful times (QED, SM) progress in particle physics did not follow
Laughlin's "emergence" catalog

Solid state physicists as Phil Anderson worry that the significant
governmental support which string theory receives maybe to the detriment of
research in their own area. From what I was able to observe the main brunt of
these regrettable developments caused by letting fundamental physics be
dominated by a TOE is string-free particle theory\footnote{At my alma mater
(the FU-Berlin), after the discontinuation of QFT the floors were filled with
people doing "ab inicio" calculations of density functionals based on the use
of appropriate computational packets.}.

For a long time physicists were critical of the suggestion that there may be a
link between the content of their science and the Zeitgeist. Indeed the
interpretation of Einstein's relativity theory in connection with the
relativism of values at the turn of the 19$^{th}$ century is a
misunderstanding caused by terminology; relativity is the theory of the
absolute i.e. the observer-independent invariants.

A similar sociological question was asked shortly after the discovery of QM.
In a book by P. Foreman \cite{Fo} the author proposes the daring thesis that a
theory in which the \textit{classical certainty} is replace by \textit{quantum
probability} could only have been discovered in war-ridden Germany where
Spengler's book \textit{the decline of the west,} which represented the post
world war I Zeitgeist in Europe, had its strongest impact. I am very sceptical
of Foreman's arguments, I think the more palatable explanation is that the
high level of German science especially on theoretical subjects was not at all
affected by the destruction of the war; intellectual life remained unaffected,
in particular it did not suffer from any state-sponsored anti-semitism.

But for the case at hand I am convinced that the building of the string theory
community and its hegemonic power on particle physics is coupled to the
millennium Zeitgeist. My arguments are as follows. After the end of the cold
war, ideologists proclaimed that global capitalism is the remaining final
social order, \textit{the end of history }\cite{Fuku}. The future world was
predicted to be democratic and everybody should expect a happy life free of
wars and social conflicts. This ideological setting turned out to be quite
successful for strengthening the hegemonic grip of globalized capitalism on
the world order.

Science is a very important part for the presentation of its power and glory,
and a theory of everything for quantum matter at the end of the millennium and
the promise of a glorious closure of fundamental physics fell on extremely
fertile ground. Superstring theory enjoys strong support in the US; and the
European Union together with other states is spending billions of dollars on
the LHC accelerator and its five detectors which among other things are
designed for the task of finding traces for two of string theories
"predictions" namely supersymmetry and extra dimensions. TV series on string
theory as \cite{nova} are unthinkable without the embedding into the
millennium's power and glory Zeitgeist.

It is quite inconceivable, that metaphoric ideas without experimental support
would have been compatible with another Zeitgeist. It is impossible that an
idea of extra dimension for which there is not the slightest indication from
particle physics and which is a graft from the already metaphoric string
setting would have been accepted at any previous time.

For a few philosophers and sociologists this surprising regressive development
did not come unexpected, especially those of the \textit{Frankfurt school of
critical theory} who anticipated a dialectic change from enlightenment into
irrationality in the development of society. According to
Horkheimer\footnote{In Horkheimer's words: \textquotedblleft!f by
enlightenment and intellectual progress we mean the freeing of man from
superstitious belief in evil forces, in demons and fairies, in blind fate --
in short, the from fear -- then denunciation of what is currently called
reason is the greatest service reason can render." cited in M. Jay, The
Dialectical Imagination. A History of the Frankfurt School and the Institute
of Social Research, 1923-1950, Univ. of California Pr., 1996, p. 253.} and
Adorno: \textit{enlightenment must convert into mythology}. Indeed the
metaphoric nature of the scientific discourse which gained acceptability
through string theory, is the ideal projection screen of a mystical
\textit{end of time beliefs at the turn of the millennium}. No other idea
coming from science had such a profound impact on the media and on popular
culture. Physics departments at renown universities \ have become the home for
a new type of scientist who spends most of her/his time travelling in order to
spread the message of extra dimensions, landscapes of multiverses etc. This
had the effect that people outside of science think of intergalactic journeys,
star wars, ufos, poltergeists (from extra dimensions) etc. when they hear the
word "superstring" \cite{Kaku}.

The metaphoric spell of string theory over particle physics cannot be
overestimated. Time dependent scattering theory, once one of the high points
of a perfect match between applied and structural aspects of QFT has been
banalized into a cooking recipe with no reliable knowledge about the resulting properties.

The corrosive influence of string theoretic metaphors have left their mark
even in new textbooks on QFT, were the derivation of time-dependent scattering
theory, once the high point of the extremely subtle particle-field relation,
has been reduced to a mere recipe, thus copying the recipe style of string
theory (where this style is the only possible one). Meanwhile important
structural insights attained in a preelectronic era vanished from the
collectice post-electronic memory. In certain cases they have been re-invented
but their remake version remained generally below the conceptual level of the
original. It may be interesting to the reader to back this up by the two
following concrete illustrations.

The structural problems behind recent proposal of unparticles have been
investigated starting from a study of the Kallen-Lehmann two-point function in
as far back as 1963 \cite{1963}. After the difficulties with incurable
infrared problems one encountered if one applied the Wigner particle picture
and (LSZ) scattering theory to electrically charged states\footnote{The
perturbative manifestation is the appearance of infrared divergencies which
can be controlled by passing to inclusive cross sections in which the
auxiliary infrared cutoffs of "virtual" photons compensate with the infrared
behavior from summation over real photons in cross sections.
Non-perturbatively the LSZ in/out fields vanish because unitarity forces the
$\delta$- mass shell singularity to spread and become softer.} it became clear
that the only way out was to look for a completely new particle concept. Since
the notion of infraparticles is fundamental for the understanding of
electrically charges states and their generating charged fields (which, as
everybody knows, are not identical with the formal Dirac fields in the
standard gauge setting of QED). An important step was the realization that the
delocalization of infraparticles is closely related to Gauss law \cite{Bu-82}
and a spontaneous breaking of Lorentz-invariance.

The theory of infraparticles is an important post renormalization contribution
to QED. The state of affairs up to the beginning of the century can be seen by
following the list of references in \cite{Por}. The main motivation namely to
overcome the inadequacy of the mass-gap based particle concept is the same for
infraparticles and its recent "unparticle" \cite{un}\cite{Grin} avatar. But
the irony is that the somewhat delocalized infraparticles are the most visible
objects in QFT; the way to come to dark matter within QFT is certainly not via
unparticles. One needs a much more radical delocalization \cite{infra}.
Phrases as: \textit{the} \textit{known particle physics is based on theories
which have a mass gap or are free in the infrared} which appear as the opening
mantra in one of the cited papers are plainly contradicted by the structure of
QED. The important property is the strength of the interaction in the infrared
region; QED has the necessary strength, whereas e.g. a pseudoscalar
interaction of nucleons with zero mass pions stays below the critical strength
in the infrared and therefore remains within the standard particle setting.

In addition to couplings between massive and massless scalar invariant matter
there is also the intimately related question about how to extract particle
informations from higher dimensional interacting (appearance of fields with
anomalous dimensions) conformal field theories. During the reign of dispersion
theory within the S-matrix setting conformal field theories fell into disgrace
since there was the suspicion that only free conformal theories are consistent
with scattering. This was put on solid mathematical grounds in
\cite{conformal} where it also argued that, since the scale invariance demands
that all multiparticle threshholds fall on top of each other, one should
expect that information on highly inclusive processes is retained. The still
open problem is whether the inclusiveness demands the summation over only
infinitely many outgoing configurations or if also an averaging over incoming
ones is required. Again the reader may want to compare this with what is
presently coming from the unparticle community.

The Araki-Haag theory of particle counters which played an important role in
the infraparticle investigations and also as models of counters for Unruh and
black hole Hawking radiation suffered a similar fate, this is at least the
impression one gets from looking at \cite{Costa}. Araki-Haag counters are
defined in terms of a subalgebra of the algebra of the so-called
\textit{almost local observables} which has the property of insensitivity with
respect to the vacuum\footnote{In the relativistic quantum mechanical DPI
scheme of section 3 the vacuum insensitivity requirement poses no problem, but
in QFT the response to the vacuum of a strictly localized counter poses a
challenge. The conclusions of the authors (who cast doubts on the Unruh
effect) apply to the DPI of section 3 but certainly not to QFT.}. These subtle
conceptual differences between a relativisic QM (the DPI in section 3) and the
omnipresence of vacuum polarizations with all the consequences for
\textit{localization thermality} in particular for a universal QFT area law
for \textit{localization entropy} with a $\ln\varepsilon$ increase in the
attenuation length $\varepsilon$ of the vacuum polarisation cloud around the
horizon \cite{interface}\cite{Hol}.

The worrisome aspect for the development of particle theory of such
re-discoveries which willfully ignore the past in order to receive the maximum
amount of attention in the present is obvious, it hampers progress and (if
noticed) creates cynics. It is our aim in this essay to highlight these
developments. In the case at hand, the early research on infraparticles is not
work on a particle physics sideline. It is \textit{the core property of any
electrically charged particle} and one may ask the question whether it is in
the nature of the turbo particle-particle physics with its instant formation
of globalized communities around any metaphoric belch of somebody with a high
viewing rate to lead to such Un-particle-physics. Hence the real blame goes to
those Journals (presumably their editors) who suffer from a disequilibrium
between high viewer ratings (and the ensuing high impact indices) but only
command over low quality referees.

To learn more about this problem let us briefly look at one of the biggest
goof-ups in history of particle theory. It was the publishing of a wrong
article in PRL \cite{Heger} claiming that Fermi's famous Gedankenexperiment
showing that QED leads to the same finite propagations speed $c$ as the
classical theory is incorrect. The author used the fact the Fermi was not
mathematically precise in order to impose his interpretation of localization
and propagation (the Born-Newton-Wigner localization) which he had elaborated
in many previous articles. This was certainly not what Fermi had in mind. The
fact that his article was published in Phys. Rev. Lett., a highly reputable
journal, had the awkward consequence that the (at that time) editor of
\textit{Nature} wrote an enthusiastic article which raised the feasibility of
time machines on the basis of these new "sensational" results. As a
consequence the subject was immediately taken up by almost all prominent daily
journals in in Europe and the US including of course the NYT.

It was not too difficult to convince the editor to accept a counter article
\cite{Bu-Yn} which rectifies the incorrect implementation of the problem. The
PRL at that time disposed over excellent referees in particle physics so this
affair was probably not caused by an uninformed referee but rather by one who
thought that a controversial article in the prestigious PRL is a good starting
point for having an academic discussion about this rather subtle issue. I
don't think that he expected an explosion into an international media hype on
time machines.

Whereas in the 90s journals still had access to referees with a broad
background extending well into the beginnings of renormalization theory, this
is not the case anymore, and an editorial board which is capable to accept and
correct its misjudgement would not fit into the present Zeitgeist. Using the
fashionable metaphors and making the right mix with interspersing noncommittal
but nice-sounding expectations about observational consequences usually
guaranties journal acceptance, whereas the publication of results with a more
mathematically and conceptually demanding conceptual background which
overstrain its post standard model readers is rejected, especially if in
addition the author is not part of the globalized particle physics community.

Again it may be helpful to back up these claims by a concrete illustration.
\ The point I want to illustrate is that vague metaphoric claims about
observable consequences are in and precise statements are out. In my arguments
linking "darkness" of matter to noncompact localization of certain objects (in
a theory which also may contain pointlike localizable subobjects) it is an
inescapable conclusion that there can be no pair creation of such dark matter
from ordinary matter \cite{dark} so if the planned experiment shows pair
creation, my proposal is out. I could not see any clear conclusion in
\cite{un}, the style of arguing is phenomenological but if it comes to the
results one finds the phrase: my goal here is not to do serious phenomenology.
How can a journal that has the highest viewing/impact rating work with such
uninformed referees who do not know that the issue of un/infra-particles was a
much researched topic more than 2 decades before to which the elite of
particle physicists e.g. people as Buchholz \cite{Bu-82} and Froehlich
\cite{Fro} contributed? Every charge particle is an unparticle in the sense of
the Kallen-Lehmann spectral function \cite{un} and certainly such particle
form the backbone of standard, visible particle physics.

It is remarkable that in the rejection of my paper which was not even allowed
to go to table of a referee it says \ "We have considered your manuscript and
conclude that it is not suited for Physical Review Letters. We make no
judgment on the correctness or technical aspects of your work. However, from
our understanding of the paper's physics results, context, and motivation, we
conclude that your paper does not have the importance and broad interest
needed for publication in our journal." Surely the metaphorical embedding was
missing, a zero production cross section it too blunt and lacks any
entertaining quality.

So my advice to any young physicist who still has to enlarge his impact
measure with the journals of highest viewer ratings is to emulate the style of
\cite{un}. Since my main purpose was to confirm my suspicions, I got
everything I wanted (although I should admit in all honesty that I probably
would not have withdrawn my contribution in case of its acceptance). To
confirm my conclusions I also submitted a paper to another high viewer rating
journal namely to JHEP about which I know from some of my colleagues that they
encountered problems. This time the difficulty started already with the
submission, since the content of my paper did not fit any of the required
subject headings. Most of them are about social constructs as D-branes, dS
vacua in string theory, intersecting branes models, F- and M-theory, (Alice in
the wonderland of the little curled) extra dimensions,....which deserve this
characterization from the observational as well as from the mathematical side.
There are veteran constructs as supersymmetry which wait for more than 4
decades to receive the observable kiss from the prince and there are titles
like AdS-CFT correspondence which one could embrace if there would not be the
uneasy feeling that in the JHEP menu they mean something metaphoric. There is
absolutely nothing on such topics as charges particles and their delocalation,
old but important issues which have not been completely solved. Since my
proposal involves selfinteracting spin 1 particles I took gauge theory and
received the following rejection.

"The interesting papers by Dr. Schroer do not belong to the areas covered by
JHEP, which is confirmed by his difficulties in finding the appropriate
keywords, as he himself admits. In view of their strong mathematical
orientation, a journal like Commun. Math. Phys. A journal on the foundation of
quantum physics (eg Foundation of Physics) could also be envisioned."

This is a silly proposal since my paper has nothing to do with mathematical
physics nor with what is commonly understood under "foundation of physics". It
is false labelling to name a journal with such a restrictive list a journal of
high energy physics. This episode points to a real worrying problem which I
have suspected behind the editorial policy of JHEP for a long time: the choice
between publishing about a subject developing from the intrinsic logic of
known physics and one which comes from social constructs of some physicists is
not any more available (there is no category "miscellaneous"). Flagships of
this new brand of particle physics as JEHP have already institutionalized the
metaphoric style.

These examples underline the main theses of the present essay; particle
physics is in a deep crisis if not to say it is in a free fall (and that only
those who are in the same reference system are not able to see this). Even if
string theory would disappear overnight, the crisis which had been growing
with it would not disappear. The content of particle theory has been pressed
into social constructs which have nothing to do with the physics nature and in
many cases do not even enjoy any mathematical control. Since this situation
has never existed before, one wonders if and how such complex peronalities as
Einstein, Dirac, or Heisenberg would operate in such an environment.

The best metaphor for the present situation which comes to my mind is that of
Laokoon fighting the snakes which try to strangle him. These snakes bear the
names of all the social constructs in particle theory. The expectation/hope
that the LHC may solve all these problems and get rid of them appears utterly
naive. How can a machine resolve a messy theoretical situation which does not
make credible predictions? Does anybody seriously believe that a veteran
social construct as supersymmetry will fade away with a whimper? And is it
possible to re-educate the participants in a 40 year march in the wrong
direction in a couple of years as e.g. the German's were re-educated after
world war II?

What is more probable is that an increasing number of people will realize that
the promised millennium power and glory was a fake; the money going into
particle physics will be cut, the number of particle physicists will go down
and if everything goes well there will be a new beginning in which
archeological study of pre-electronic work will play an important role. It
will be a world in which the solution of important unsolved problems is as
important as the pursuit of new discoveries. It certainly will not be a world
in which one is forced to follow social constructs which are determined by
viewer rates.

The title of this last section is a reminder of the futility of looking for
individual scapegoats in connection with the ongoing crisis. Whereas it is
true that the uncritical support of prominent theorists has contributed to the
present situation, it is equally correct to add that this would not have been
possible without the tacit approval of all of us. We all (i.e. we particle
theoreticians) have tacitly accepted its growth and some of us have even
linked their fortune to it.

The conquest of QED, the first renormalized QFT, was not possible without a
major conceptual investment. In comparison the discovery of the standard model
was rather straightforward from a conceptual viewpoint; the required
improvements were related to renormalization technology of selfinteracting
massive s=1 fields; This (apart from the observational verifications)
relatively easy conquest\footnote{According to our best knowledge a
renormalizable selfinteracting massive s=1 particles must be accompanied by a
massive scalar particle i.e. one does not have to discover that particluar s=1
interaction which enters the standard model because there is only one
interaction.} may have given rise to the somewhat presumptuous state of mind
which led to the idea of a TOE, thus re-creating that mixture of blindness and
intellectual arrogance which already accompanied pre-string proposals of a TOE
or Weltformel.

A invariable by-product of TOEs is that they, like ideologies and religious
beliefs, have a tendency to attract mentally unbalanced people. As a result of
the highly abstract nature of the subject these are people who have a problem
not with their intellectual capacity; rather their problem is with the
malfunctions of their emotions in that they are unable to control their
aggressive behavior against others who do not share their opinions. It is a
known fact that e.g. statistically the frequency of Asperger's syndrome is
higher in the community of theoretical physicists and mathematicians than in
other professions..

A notorious and quite public case is that of a junior Harvard University
professor who drove the abuse of fellow physicists (including myself) to
unprecedented heights by calling them all kind of names in public blogs which
serve as a critical forum for string theory and related subjects. When this
individual was using death threats against a fellow physicists I thought it is
time to quit my participation and I also decided to rewrite this essay which
contained many reactions to the comments of this individual which after this
episode did not any longer make sense. I thought it was more appropriate to
criticize the scientific content of string theory than to comment on the
statements of an obviously emotionally disordered individual. No area of human
activity is immune against attracting individuals like this and hence string
theory should be criticized on its scientific merits and not on the basis of
an epiphenomenon.

What really shocked me was not the bizarre behavior of one pro-string
individual. Rather It was the silence of his colleagues from the Harvard
string/extra dimension group for whom he obviously served as a useful fighter
for their course. An episode like this is not covered by the title of this
last section.

\begin{acknowledgement}
I thank Fritz Coester for sharing with me his recollection about the
Stueckelberg-Heisenberg S-matrix dispute. I also recollect with pleasure
several encounters I had with a young string theorist named Oswaldo Zapata
whose growing critical attitude and disenchantment with this area of research
led him to the previous version of this article and away from string theory
into the philosophy, sociology and history of science.
\end{acknowledgement}

\section{An epilog, reminiscences about encounters with Juergen Ehlers}

Shortly after my this year arrival in Berlin I received the sad news about
Juergen Ehlers sudden death. As every year, I was looking forward to meet him
at the AEI in Golm. Being roughly from the same generation (I am only 4 years
younger than Juergen). Having been a member of Pascual Jordan's relativity
group (before I decided at the beginning of DESY to continue in particle
physics under Harry Lehmann's guidance), we naturally had a similar
motivational and philosophical background. Juergen was much more advanced and
he together with Engelbert Schuecking (now a retired professor at the NYU)
were my role models.

The belief that the delicate equilibrium between speculative attempts at
innovation and their critical review is crucial in fundamental physics and the
uneasy feeling that we are presently witnessing a derailment towards an
unhealthy metaphoric unrestrained speculative side was certainly shared by
Juergen. His increasing interest in the foundations of quantum theory and his
appreciation of the deep conceptual differences between quantum mechanics and
quantum field theory which I supported with my knowledge of local quantum
physics was the motor of most of our discussions in his office at the AEI in
Golm near Berlin.

Part of this interest originated from Juergen's interest to understand
Jordan's role as the protagonist of QFT i.e. the physics content of Sam
Schweber's poetic words: "Jordan is the unsung hero of QFT" and why Dirac, who
favored for a long time a very successful quantum mechanical particle based
setting\footnote{Historically the notion of antiparticles originated through
Dirac's hole theory and not through field quantization.}, finally, at the
beginning of the 50's, came around to adopt the more radical quantum field
setting beyond electromagnetic radiation for all kind of matter and radiation.

In 2004 there was a conference about Jordan's contributions to the foundations
of quantum physics where many of the results lost in time (a collateral damage
of the great political upheavals) were reviewed and presented to an astonished
audience. While scanning the archives of the CBPF library in Rio de Janeiro
(to where I moved after my retirement) I came across a 1929 status report
about QFT by Pascual Jordan which was quite a revelation to me. In this report
Jordan was pleading for a new QFT which is not based on a classical
parallelism. Purely on philosophical grounds he was distancing himself from
his only 4 year old brainchild "die Quantelung der Wellenfelder". The reason
why this plea of Jordan resonated with me was that ever since my PhD under
Lehmann I was working on an autonomous setting of QFT (algebraic QFT, local
quantum physics), a program initiated by Rudolf Haag almost 30 years after
Jordan's plea in his 1929 plenary talk. In view of the total absence of
conceptual-mathematical instruments and concepts this spirited plea was truly
amazing. After I made these connections in my talk at that conference
dedicated to the memory of Jordan and his work Juergen became very interested
to learn something about the present status of this program.

At that time Juergen also asked me about my opinion on Jordan's algebraic
construction of magnetic monopoles. Shortly before I had seen a purely
algebraic derivation by Roman Jackiew on the arXiv. I wrote to Roman and he
was probably as surprised as I to find the full argument with all details in
Jordan's 3 page paper (whose 1934 publication in Zeitschrift fuer Physik "was
a burial of first class", a phrase attributed to the relativist P. Bergmann
about publications in Z.f.Ph. during the third Reich).

In Jordan's work on what is nowadays referred to \textit{Transformation
Theory} (the equivalence of the different formulations of QT) he thanks Fritz
London for sending his results on this issue before publication and strongly
praises the author for the clarity of his presentation. Whereas I took this as
matter of encouragement of a young novice to QT in accordance with the more
polite social etiquette of the times, Juergen went to the library and red
London's article. He convinced me that, apart from any politeness, Jordan
really had profound scientific reasons to be impressed. London's article is
the first article which connects the different formulations with the
appropriate mathematical tools as the concept of Hilbert space and rotations
therein (unitary operators). Usually one attributes the implementation of
these mathematical notions into QM to John von Neumann and overlooks Fritz
London, who wrote this impressive article when he was an assistant at the
Technische Hochschule in Stuttgart i.e. when he certainly was not part of the
great quantum dialog which started between Goettingen, Copenhagen and Hamburg
and spread to other places after London's publication.

After the Mainz conference \cite{Mainz} on the occasion of Jordan's 100
birthday, Juergen was engaged in a book project about Jordan and in this
context he asked me for some advice. In particular he wanted to know whether
Jordan's several papers on what he called the "neutrino theory of light"
should also enter such a project. Since Jordan's contemporaries made fun of
this pet idea of his\footnote{The critical reaction against the metaphoric
"neutrino theory of light" title of Jordan's papers caused his contemporaries
to overlook their very interesting bosonization/fermionization content.} (in
\cite{Pais} one even finds a very funny mocking song), Juergen was in favor of
ignoring all of those articles about this topic which looked to him (and even
more so to Jordan's contemporaries very suspicious). I finally convinced
Juergen not to do this. The series of articles under the title the
\textit{neutrino theory of light} have of course nothing to do with neutrinos
nor with light; in those articles Jordan discovered what is nowadays called
the bosonization of fermions (or fermionization of bosons) which is a typical
structural property of 2-dim. conformal theories.

Jordan obviously saw the potential relevance of this property, but unlike
Luttinger almost 3 decades later, solid state physics (where low dimensional
QFTs in certain circumstances turn out to be very useful) had yet no need for
such observations on QFT structures. In order to "sell" his nice field
theoretic result he used the very metaphoric "neutrino theory of light" label.
The false labeling backfired, the result was not only the mentioned mockery by
his contemporaries, but also the fact that when bosonozation really became an
interesting issue in the 60's, nobody expected a the intersteing content about
bosonization behind Jordan's metaphoric title; \textit{such things simply did
not work well in those times}.

Nowadays papers in which not only the title but also the content is metaphoric
have the best chance to be accepted in career supporting journals especially
if their content fits with the socially accepted metaphors of their subject
lists and in this way they cause the leat amount of work to fashion-conscious
referees. I reminded Juergen of the actual world we live in, where the
combination of metaphoric title with an interesting rather well-defined
non-metaphoric content is rare. As a result he told me that he will retain at
least one paper of the neutino theory of light topic from the dawn of
bosonization/fermionization. Obviously Juergen realized that in view of the
present crisis in particle physics the metaphoric faux pas of Jordan, for
which he was mocked by his contemporaries, was a nostalgic memory about better
times. With this remark we have returned to main theme of this essay and its
wistful contrast with a style of physics of which is so excellently conveyed
in the person of Juergen Ehlers

With all differences in the interpretation of Jordan's legacy removed, I was
looking forward to new encounter with Juergen in order to enlarge my feeble
knowledge on matters of astrophysics and general relativity (gravitational
lensing,..) in order to broaden the basis of my recent interest in dark
matter. As a "payback" I wanted to explain to Juergen in more detail the
fundamental conceptual differences between QFT and relativistic QM concerning
the totally different forms of "entanglement". My aim was to convince him that
the black hole information loss (a 40 year old problem which led to a still
ongoing controversy) is a pseudo-problem which resulted from confusing the
information-based ("Gedanken") entanglement coming from quantum mechanical
Born-localization with the kind of entanglement coming from field theoretic
localization. The latter leads to the well-known thermal manifestation
(including a universal area proportionality which has a pure local quantum
physical origin independent of the classical area law which Bekenstein derived
in Einstein-Hilbert like classical field theories). This time I was much
better prepared as last year since I had done a lot of homework
\cite{interface}.

I miss Juergen, I cannot think of anybody as knowledgable, as interested in
fundamental questions beyond his own area and at the same time endowed with an
amazing critical power.

\bigskip

\end{document}